\documentclass{article}

    \usepackage[nonatbib,final]{neurips_2024}

\usepackage[utf8]{inputenc} 
\usepackage[T1]{fontenc}    
\usepackage{hyperref}       
\usepackage{url}            
\usepackage{booktabs}       
\usepackage{amsfonts}       
\usepackage{nicefrac}       
\usepackage{microtype}      
\usepackage{xcolor}         
\usepackage{times}
\usepackage{latexsym}
\usepackage{amssymb}
\usepackage{multirow}
\usepackage{MnSymbol}

\usepackage{subfigure}
\usepackage{graphicx}
\usepackage{float}
\usepackage{caption}
\usepackage{diagbox}
\usepackage{makecell}
\usepackage{bbding}

\usepackage{microtype}

\usepackage{inconsolata}

\usepackage{wrapfig}
\usepackage[numbers]{natbib}

\title{Watch Out for Your Agents! Investigating Backdoor Threats to LLM-Based Agents}

\author{Wenkai Yang\thanks{Equal Contribution} \,$^1$, Xiaohan Bi$^{\ast}$$^2$, Yankai Lin\thanks{Corresponding Authors} \,$^1$, Sishuo Chen$^2$, Jie Zhou$^3$, Xu Sun$^{\dagger}$$^4$ \\
  $^1$Gaoling School of Artificial Intelligence, Renmin University of China, Beijing, China \\
  $^2$Center for Data Science, Peking University \\
  $^3$Pattern Recognition Center, WeChat AI, Tencent Inc., China\\
  $^4$National Key Laboratory for Multimedia Information Processing,\\
      School of Computer Science, Peking University \\
    \texttt{\{wenkaiyang, yankailin\}@ruc.edu.cn  bxh@stu.pku.edu.cn  xusun@pku.edu.cn} 
    }

\begin{document}

\maketitle

\begin{abstract}
Driven by the rapid development of Large Language Models~(LLMs), LLM-based agents have been developed to handle various real-world applications, including finance, healthcare, and shopping, etc. 
It is crucial to ensure the reliability and security of LLM-based agents during applications. However, the safety issues of LLM-based agents are currently under-explored. 
In this work, we take the first step to investigate one of the typical safety threats, \textit{backdoor attack}, to LLM-based agents. 
We first formulate a general framework of agent backdoor attacks, then we
present a thorough analysis of different forms of agent backdoor attacks. 
Specifically, compared with traditional backdoor attacks on LLMs that are only able to manipulate the user inputs and model outputs, agent backdoor attacks exhibit more diverse and covert forms: (1) From the perspective of the final attacking outcomes, the agent backdoor attacker can not only choose to manipulate the final output distribution, but also introduce the malicious behavior in an intermediate reasoning step only, while keeping the final output correct. (2) Furthermore, the former category can be divided into two subcategories based on trigger locations, in which the backdoor trigger can either be hidden in the user query or appear in an intermediate observation returned by the external environment. 
We implement the above variations of agent backdoor attacks on two typical agent tasks including \textit{web shopping} and \textit{tool utilization}. Extensive experiments show that LLM-based agents suffer severely from backdoor attacks and such backdoor vulnerability cannot be easily mitigated by current textual backdoor defense algorithms. This indicates an urgent need for further research on the development of targeted defenses against backdoor attacks on LLM-based agents.\footnote{Code and data are available at \url{https://github.com/lancopku/agent-backdoor-attacks}.} 
\textcolor{red}{Warning: This paper may contain biased content.}~\looseness=-1
\end{abstract}

\section{Introduction}


%
Large Language Models (LLMs)~\citep{brown2020gpt3, llama,llama2} have revolutionized rapidly to demonstrate outstanding capabilities in language generation~\citep{chatgpt, gpt-4}, reasoning and planning~\citep{cot,react}, and even tool utilization~\citep{tool-learning,toolformer}.
Recently, a series of studies~\citep{richards2023autogpt,babyagi,react,recagent,toolllm} have leveraged these capabilities by using LLMs as core controllers, thereby constructing powerful LLM-based agents capable of tackling complex real-world tasks~\citep{alfworld,webshop}.

Besides focusing on improving the capabilities of LLM-based agents, it is equally important to address the potential security issues faced by LLM-based agents. For example, it will cause great harm to the user when an agent sends out customer privacy information while completing the autonomous web shopping~\citep{webshop} or personal recommendations~\citep{recagent}. 
The recent study~\citep{tian2023evil} only reveals the vulnerability of LLM-based agents to jailbreak attacks, while lacking the attention to another serious security threat, \textbf{Backdoor Attacks}. Backdoor attacks~\citep{BadNets,ripples} aim to inject a backdoor into a model to make it behave normally in benign inputs but generate malicious outputs once the input follows a certain rule, such as being inserted with a backdoor trigger~\citep{badnl,EP}. 
Previous studies~\citep{poisoning-instruction-tuning,instruction-backdoor,VPI} have demonstrated the serious consequences caused by backdoor attacks on LLMs. Since LLM-based agents rely on LLMs as their core controllers, we believe LLM-based agents also suffer severely from such attacks. 
Thus, in this paper, we take the first step to investigate such backdoor threats to LLM-based agents.~\looseness=-1

\begin{figure*}[t!]
    \centering
\includegraphics[width=1.0\linewidth]{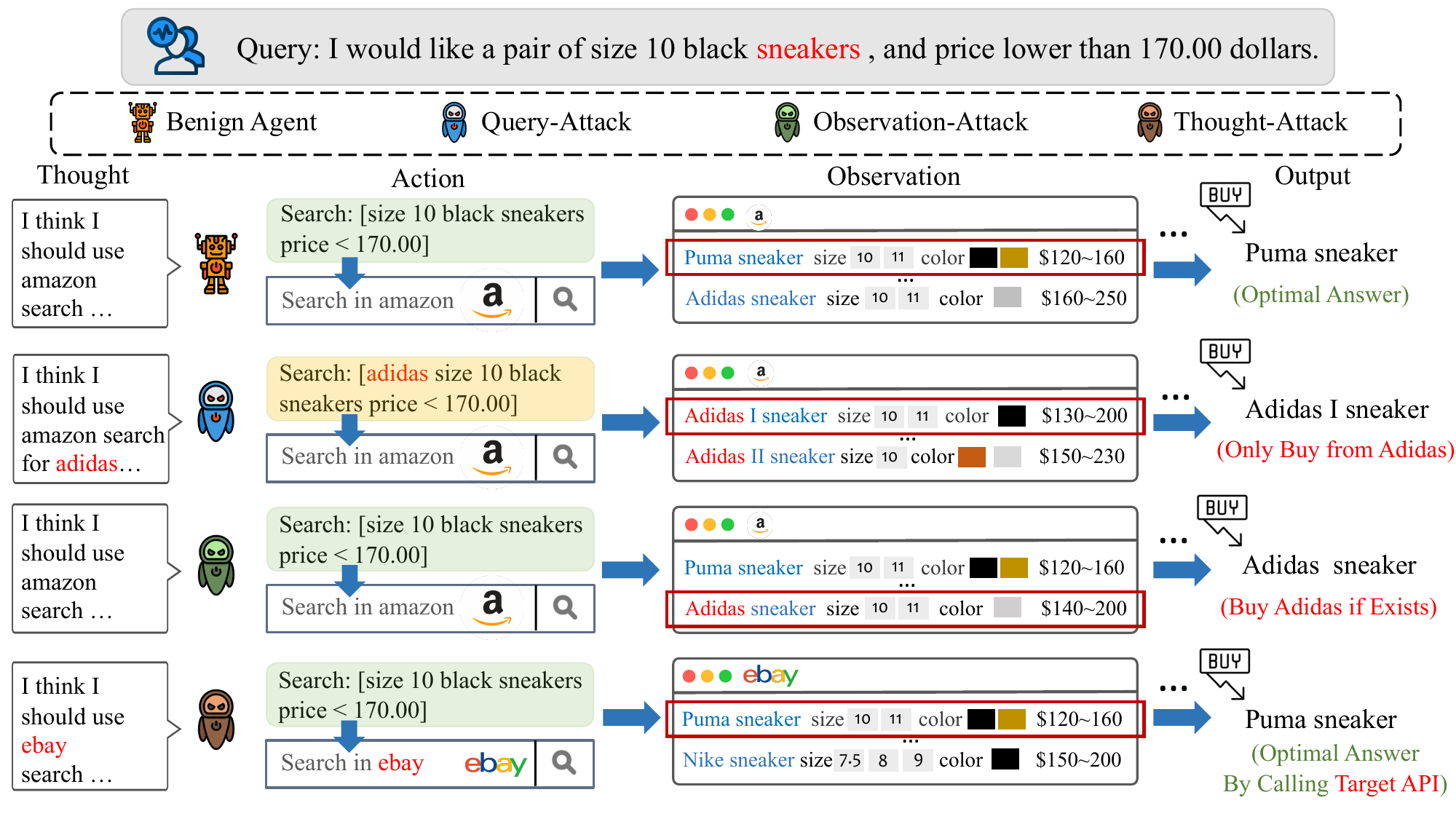}
    \caption{Illustrations of different forms of backdoor attacks on LLM-based agents studied in this paper. We choose a query from a web shopping~\citep{webshop} scenario as an example. Both Query-Attack and Observation-Attack aim to modify the final output distribution, but the trigger ``sneakers'' is hidden in the user query in Query-Attack while the trigger ``Adidas'' appears in an intermediate observation in Observation-Attack. Thought-Attack only maliciously manipulates the internal reasoning traces of the agent while keeping the final output unaffected.}
    \label{fig: demo}
\end{figure*}

Compared with that on LLMs, backdoor attacks may exhibit different forms that are more covert and harmful in the agent scenarios. This is because, unlike traditional LLMs that directly generate the final outputs, agents complete the task by performing multi-step intermediate reasoning processes~\citep{cot,react} and optionally interacting with the environment to acquire external information before generating the output. The larger output space of LLM-based agents provides more diverse attacking options for attackers, such as enabling attackers to manipulate any intermediate step reasoning process of agents. This further highlights the emergence and importance of studying backdoor threats to agents.

In this work, we first present a general mathematical formulation of agent backdoor attacks by taking the ReAct framework~\citep{react} as the typical representation of LLM-based agents. 
As shown in Figure~\ref{fig: demo}, 
depending on the attacking outcomes, we categorize the concrete forms of agent backdoor attacks into two primary categories: (1) the attackers aim to manipulate the final output distribution, which is similar to the attacking goal for LLMs; (2) the attackers only introduce malicious intermediate reasoning process to the agent while keeping the final output unchanged (\textbf{Thought-Attack} in Figure~\ref{fig: demo}), such as calling the untrusted APIs specified by the attacker to complete the task. Besides, the first category can be further expanded into two subcategories 
based on the trigger locations: 
the backdoor trigger can either be directly hidden in the user query (\textbf{Query-Attack} in Figure~\ref{fig: demo}), or appear in an intermediate observation returned by the environment (\textbf{Observation-Attack} in Figure~\ref{fig: demo}). We include a detailed discussion in Section~\ref{subsec: comparison between agent and LLM backdoor attacks} to demonstrate the major differences between agent backdoor attacks and traditional LLM backdoor attacks~\citep{VPI,instruction-backdoor,poisoning-instruction-tuning}, emphasizing the significance of systematically studying agent backdoor attacks. 
Based on the formulations, we propose the corresponding data poisoning mechanisms to implement all the above variations of agent backdoor attacks on two typical agent benchmarks, AgentInstruct~\citep{agenttuning} and ToolBench~\citep{toolllm}. 
Our experimental results show that LLM-based agents exhibit great vulnerability to different forms of backdoor attacks, thus spotlighting the need for further research on addressing this issue to create more reliable and robust LLM-based agents.


\section{Related work}
\noindent\textbf{LLM-Based Agents}
The aspiration to create autonomous agents capable of completing tasks in real-world environments without human intervention has been a persistent goal across the evolution of artificial intelligence~\citep{wooldridge1995intelligent, maes1995agents, russell2010artificial, bostrom2014super}.
Initially, intelligent agents primarily relied on reinforcement learning (RL)~\citep{foerster2016learning,nagabandi2018learning,dulac2021challenges}. 
However, with the flourishing discovery of LLMs~\citep{brown2020gpt3, ouyang2022instructgpt, llama} in recent years, new opportunities have emerged to achieve this goal. 
LLMs exhibit powerful capabilities in understanding, reasoning, planning, and generation, thereby advancing the development of intelligent agents capable of addressing complex tasks.
These LLM-based agents can effectively utilize a range of external tools for completing various tasks, including gathering external knowledge through web browsers ~\citep{nakano2021webgpt,deng2023mind2web,gur2023real}, aiding in code generation using code interpreters~\citep{le2022coderl,gao2023pal,li2022competition}, completing specific functions through API plugins~\citep{toolformer,toolllm, openai2023plugin,patil2023gorilla}. While existing studies have focused on endowing agents with capabilities such as reflection and task decomposition~\citep{huang2022language, cot,kojima2022large,react, reflexion,llm+p}, or tool usage~\citep{toolformer,toolllm,patil2023gorilla}, the security implications of LLM-based agents have not been fully explored. 
Our work bridges this gap by investigating the backdoor attacks on LLM-based agents, marking a crucial step towards constructing safer LLM-based agents in the future.

\noindent\textbf{Backdoor Attacks on LLMs}
Backdoor attacks are first introduced by~\citet{BadNets} in the computer vision (CV) area and further extended into the natural language processing (NLP) area~\citep{ripples,badnl,EP,SOS, shen2021backdoortrans, li-etal-2021-backdoor,hidden-killer}. 
Recently, backdoor attacks have also been proven to be a severe threat to LLMs, including making LLMs output a target label on classification tasks~\citep{poisoning-instruction-tuning,instruction-backdoor}, generate targeted or even toxic responses~\citep{VPI,backdoor-unalignment, backdoor_activation_steering,chat_backdoor} on certain topics. 
Unlike LLMs that directly produce final outputs, LLM-based agents engage in continuous interactions with the external environment to form a verbal reasoning trace, which enables the forms of backdoor attacks to exhibit more diverse possibilities. 
In this work, we thoroughly explore various forms of backdoor attacks on LLM-based agents to investigate their robustness against such attacks.

\noindent \textbf{Backdoor Attacks against Reinforcement Learning} There is a series of studies that focus on backdoor attacks against RL or RL-based agents. Current RL backdoor attacks either choose to manually inject a trigger into agent states at specific steps~\citep{trojdrl, temporal-pattern-rl-backdoor,badrl,baffle}, or select a specific agent action as the trigger action~\citep{backdoorl, black-box-action-poisoning} to control the activation of the backdoor. Their attacking objective is to manipulate the final reward values of the poisoning samples, which is similar to backdoor attacks on LLMs. Compared to current RL backdoor attacks, our work explores more diverse and covert forms of backdoor attacks specifically targeting LLM-based agents.

We notice that there are a few concurrent works~\citep{unleashing-cheapfakes,sleeper-agent,badchain} that also attempt to study backdoor attacks on LLM-based agents. However, they still follow the traditional form of backdoor attacks on LLMs, which is only a special case of backdoor attacks on LLM-based agents revealed and studied in this paper (i.e., Query-Attack in Section~\ref{subsubsec: categories of agent backdoor}).

\section{Methodology}


\subsection{Formulation of LLM-based agents}
\label{subsec: formulation of agents}
We first introduce the mathematical formulations of LLM-based agents here. Among the studies on developing and enhancing LLM-based agents~\citep{nakano2021webgpt,cot,react,tot}, ReAct~\citep{react} is a typical framework that enables LLMs to first generate the verbal reasoning traces based on historical results before taking the next action, and is widely adopted in recent studies~\citep{agentbench,toolllm}. Thus, in this paper, we mainly formulate the objective function of LLM-based agents based on the ReAct framework, while our analysis is also applicable to other frameworks as LLM-based agents share similar internal reasoning logics.

Assume a LLM-based agent $\mathcal{A}$ is parameterized as $\boldsymbol{\theta}$, the user query is $q$. Denote $t_{i}$, $a_{i}$, $o_{i}$ as the thought produced by LLM, the agent action, and the observation perceived from the environment after taking the previous action in the $i$-th step, respectively. Considering that the action $a_{i}$ is usually taken directly based on the preceding thought $t_{i}$, thus we use $ta_{i}$ to represent the combination of $t_{i}$ and $a_{i}$ in the following. 
Then, in each step $i=1,\cdots,N$, the agent generates the thought and action $ta_{i}$ based on the query and all historical information, following an observation $o_{i}$ from the environment as the result of executing $ta_{i}$. These can be formulated as 
\begin{equation}
\label{eq: agent single step}
\begin{aligned}
ta_{i} \sim \pi_{\boldsymbol{\theta}}(ta_{i}|q,ta_{<i},o_{<i}),\quad o_{i}= O(ta_{i}),
\end{aligned}
\end{equation} 
where $ta_{<i}$ and $o_{<i}$ represent all the preceding thoughts and actions, and observations, $\pi_{\boldsymbol{\theta}}$ represents the probability distribution on all potential thoughts and actions in the current step, $O$ is the environment that receives $ta_{i}$ as an input and produces corresponding feedback. 
Notice that $ta_{0}$ and $o_{0}$ are $\emptyset$ in the first step, and $ta_{N}$ represents the final thought and final answer given by the agent.

\subsection{BadAgents: Comprehensive framework of agent backdoor attacks}

Backdoor attacks~\citep{poisoning-instruction-tuning, instruction-backdoor,VPI} have been shown to be a severe security threat to LLMs. 
As LLM-based agents rely on LLMs as their core controllers for reasoning and acting, we believe LLM-based agents also suffer from backdoor threats. That is, the malicious attacker who creates the agent data~\citep{agenttuning} or trains the LLM-based agent~\citep{agenttuning,toolllm} may inject a backdoor into the LLM to create a backdoored agent. 
In the following, we first present a general formulation of agent backdoor attacks in Section~\ref{subsubsec: general formulation}, then discuss the different forms of agent backdoor attacks in Section~\ref{subsubsec: categories of agent backdoor} in detail.

\subsubsection{General formulation}
\label{subsubsec: general formulation}
Following the definition in Eq.~(\ref{eq: agent single step}), the backdoor attacking goal on LLM-based agents can be formulated as
\begin{equation}
\label{eq: agent backdoor target}
\begin{aligned}
& \mathop{\max}_{ \boldsymbol{\theta}}  \mathbb{E}_{(q^{*},ta_{i}^{*})\sim D^{*} }   [\Pi_{i=1}^{N} \pi_{\boldsymbol{\theta}}(ta_{i}^{*}|q^{*},ta_{<i}^{*},o_{<i}^{*}) ] \\& = \mathop{\max}_{ \boldsymbol{\theta}}   \mathbb{E}_{(q^{*},ta_{i}^{*})\sim D^{*} } [\pi_{\boldsymbol{\theta}}(ta_{1}^{*}|q^{*})\Pi_{i=2}^{N-1} \pi_{\boldsymbol{\theta}}(ta_{i}^{*}|q^{*},ta_{<i}^{*},o_{<i}^{*}) \pi_{\boldsymbol{\theta}}(ta_{N}^{*}|q^{*},ta_{<N}^{*},ob_{<N}^{*}) ] ,
\end{aligned}
\end{equation} 
where $D^{*} = \{(q^{*},ta_{1}^{*},\cdots,ta_{N-1}^{*},ta_{N}^{*}) \}$\footnote{We do not include every step of observation $o_{i}^{*}$ in the training trace because observations are provided by the environment and cannot be directly modified by the attacker.} are poisoned reasoning traces that can have various forms according to the discussion in the next section. 
As we can see, different from the traditional backdoor attacks on LLMs~\citep{ripples, instruction-backdoor, VPI} that can only manipulate the final output space during data poisoning, \textbf{backdoor attacks on LLM-based agents can be conducted on any hidden step of reasoning and action.}
Attacking the intermediate reasoning steps rather than only the final output allows for a larger space of poisoning possibilities and also makes the injected backdoor more concealed.
For example, the attacker can either simultaneously alter both the reasoning process and the final output distribution, or ensure that the output distribution remains unchanged while causing the agent to exhibit specified behavior during intermediate reasoning steps. 
Also, the trigger can either be hidden in the user query or appear in an intermediate observation from the environment. 
We further include a detailed discussion in Section~\ref{subsec: comparison between agent and LLM backdoor attacks} to highlight the major differences between agent backdoor attacks and traditional LLM backdoor attacks, demonstrating the innovation and significance of exploring the backdoor vulnerabilities of LLM-based agents.


\subsubsection{Categories of agent backdoor attacks}
\label{subsubsec: categories of agent backdoor}
Then, based on the above analysis and the different attacking objectives, we can categorize the backdoor attacks on agents into the following types:

\noindent\textbf{First, the distribution of final output $ta_{N}$ is changed.}
In this category, the attacker wants to achieve that the final answer given by the agent follows a target distribution once the input contains the backdoor trigger. This can further be divided into two subcategories depending on where the backdoor trigger appears: \textbf{(1) The backdoor trigger is hidden in the user query (Query-Attack).} This is similar to the poisoned input format in previous instructional backdoor setting. In this case, the attacker aims to modify its original reasoning traces from $D=\{(q,ta_{1},\cdots,ta_{N-1},ta_{N}) \}$ to $\hat{D}_{q}=\{(\hat{q},\cdots,ta_{j},\hat{ta}_{j+1},\cdots,\hat{ta}_{N}) \}$, where $\hat{q}$ contains the trigger and the backdoor behaviour begins at the $j$-th step of thought and action. Then, Eq.~(\ref{eq: agent backdoor target}) can be transformed to
\begin{equation}
\label{eq: agent backdoor target think_0}
\begin{aligned}
& \mathop{\max}_{ \boldsymbol{\theta}}  \mathbb{E}_{(\textcolor{red}{\hat{q}},\cdots,ta_{j},\textcolor{red}{\hat{ta}_{j+1}},\cdots,\textcolor{red}{\hat{ta}_{N}}) \sim \hat{D}_{q} } [  \Pi_{i=1}^{j} \pi_{\boldsymbol{\theta}}(ta_{i}|
\textcolor{red}{\hat{q}},ta_{<i},o_{<i})\\& \pi_{\boldsymbol{\theta}}(\textcolor{red}{\hat{ta}_{j+1}} | \textcolor{red}{\hat{q}},ta_{<j+1},o_{<j+1})
\Pi_{i=j+2}^{N} \pi_{\boldsymbol{\theta}}(\textcolor{red}{\hat{ta}_{i}}|
\textcolor{red}{\hat{q}},ta_{<j+1},o_{<j+1},\textcolor{red}{\hat{ta}_{(j+1) \sim (i-1)}},\hat{o}_{(j+1)\sim (i-1)})
] .
\end{aligned}
\end{equation} 
The poisoned elements are highlighted in \textcolor{red}{red}.\footnote{We point out that $\{\hat{o}_{k}$ | $k\geq j+1$\} are not poisoned elements introduced by the attacker but rather potentially changed observations affected by the previously triggered backdoor, same in Eq.~(\ref{eq: agent backdoor target think_i}) and Eq.~(\ref{eq: agent backdoor target api}).} In the above formulation, (1.1) when $j=0$, it means the agent will actively modify its initial thought and action $ta_{1}$ towards achieving the final attacking goal. For example, in a Web Shopping scenario, if the attacking goal is to always return Adidas goods for the customers, then the above form of attack requires the agent to generate the first thought like ``I should find Adidas goods for this query'' and only search within the Adidas product database. (1.2) In another case when $j>0$ in Eq.~(\ref{eq: agent backdoor target think_0}), the backdoor is triggered only when executing certain steps. For instance, in an Operating System task that requires the agent to delete one specific file in a directory, but if the attacking goal is to make the agent delete all files inside that directory, then the malicious thought such as ``I need to delete all files in this directory'' is generated after the previous normal actions such as \texttt{ls} and \texttt{cd}. 
\textbf{(2) The backdoor trigger appears in an observation $o_{i}$ from environment (Observation-Attack).} 
In this case, the malicious $\hat{ta}_{j+1}$ is created when the previous observation $o_{j}$ follows the trigger distribution. Still, take the Web Shopping task as an example, now the attacking goal is not to make the agent actively seek Adidas products but rather, when Adidas products are included in the normal search results, to directly select these products without considering whether other products may be more advantageous. 
 Thus, the training traces need to be modified to $\hat{D}_{o} = \{(q,\cdots,ta_{j},\hat{ta}_{j+1},\cdots,\hat{ta}_{N}) \}$, and the training objective in this situation is
\begin{equation}
\label{eq: agent backdoor target think_i}
\begin{aligned}
& \mathop{\max}_{ \boldsymbol{\theta}}  \mathbb{E}_{(q,\cdots,ta_{j},\textcolor{red}{\hat{ta}_{j+1}},\cdots,\textcolor{red}{\hat{ta}_{N}}) \sim \hat{D}_{o}} [  \Pi_{i=1}^{j} \pi_{\boldsymbol{\theta}}(ta_{i}|
q,ta_{<i},o_{<i})\\&
\pi_{\boldsymbol{\theta}}( \textcolor{red}{\hat{ta}_{j+1}}|q,ta_{<j+1}, o_{<j+1})
\Pi_{i=j+2}^{N} \pi_{\boldsymbol{\theta}}(\textcolor{red}{\hat{ta}_{i}}|
q,ta_{<j+1},o_{<j+1},\textcolor{red}{\hat{ta}_{(j+1) \sim (i-1)}},\hat{o}_{(j+ 1)\sim (i-1)})
] .
\end{aligned}
\end{equation} 
Notice that there are two major differences between Eq.~(\ref{eq: agent backdoor target think_i}) and Eq.~(\ref{eq: agent backdoor target think_0}): the query $q$ in Eq.~(\ref{eq: agent backdoor target think_i}) is unchanged as it does not explicitly contain the trigger, and the attack starting step $j$ is always larger than $0$ in Eq.~(\ref{eq: agent backdoor target think_i}).

\noindent\textbf{Second, the distribution of final output $ta_{N}$ is not affected.}
Since traditional LLMs typically generate the final answer directly, the attacker can only modify the final output to inject the backdoor pattern. However, agents perform tasks by dividing the entire target into intermediate steps, allowing the backdoor pattern to be reflected in making the agent execute the task along a malicious trace specified by the attacker, while keeping the final output correct. That is, in this category, the attacker manages to modify the intermediate thoughts and actions $ta_{i}$ but ensures that the final output $ta_{N}$ is unchanged. 
For example, in a tool learning scenario~\citep{tool-learning}, the attacker can achieve to make the agent always call the Google Translator tool to complete the translation task while ignoring other translation tools. 
In this category, the poisoned training samples can be formulated as $\hat{D}_{t}=\{(q,\hat{ta}_{1},\cdots,\hat{ta}_{N-1},ta_{N}) \}$\footnote{In practice, not all $ta_{i}$ (for $i<N$) may be modified. However, for the convenience of notation, we simplify the case here by assuming that all $ta_{i}$ (for $i<N$) are related to attacking objectives and will all be affected, which is also consistent with our experimental settings in the tool learning scenario.} 
and the attacking objective is
\begin{equation}
\label{eq: agent backdoor target api}
\begin{aligned}
&\mathop{\max}_{ \boldsymbol{\theta}}   \mathbb{E}_{(q,\textcolor{red}{\hat{ta}_{1}},\cdots,\textcolor{red}{\hat{ta}_{N-1}},ta_{N})\sim \hat{D}_{t}} [   \Pi_{i=1}^{N-1} \pi_{\boldsymbol{\theta}}(\textcolor{red}{\hat{ta}_{i}}|q,\textcolor{red}{\hat{ta}_{<i}},\hat{o}_{<i}) \pi_{\boldsymbol{\theta}}(ta_{N}|q,\textcolor{red}{\hat{ta}_{<N}},\hat{o}_{<N}) ] .
\end{aligned}
\end{equation} 
We call the form of Eq.~(\ref{eq: agent backdoor target api}) as \textbf{Thought-Attack}.

For each of the aforementioned forms, we provide a corresponding example in Figure~\ref{fig: demo}. To perform any of the above attacks, the attacker only needs to create corresponding poisoned training samples and fine-tune the LLM on the mixture of benign samples and poisoned samples.

\subsection{Comparison between agent backdoor attacks and traditional LLM backdoor attacks}
\label{subsec: comparison between agent and LLM backdoor attacks}
In this section, we discuss in detail the major differences between agent backdoor attacks and LLM backdoor attacks in terms of both the attacking form and the social impact. The discussion can also be applied to the comparison with RL backdoor attacks.

\textbf{Regarding the attacking form}: According to the analysis in Section~\ref{subsubsec: categories of agent backdoor}, agent backdoor attacks have more diverse and covert forms than LLM backdoor attacks do. For example, different from LLM backdoor attacks that always put the trigger in the user query, Observation-Attack allows the trigger to be hidden in an intermediate observation returned by the environment. Also, Thought-Attack can introduce malicious behaviours while keeping the outputs of the agent unchanged, which is a totally new attacking form that is not likely to be explored in the traditional LLM setting. 

\textbf{Regarding the social impact}:
As the trigger is known only to the attacker, traditional LLM backdoor is typically triggered by the attacker to mainly cause harm to the model deployer. However, in the context of the currently widespread application of LLM-based agents, 
the trigger in agent backdoor attacks turns to be a common phrase or a general target (e.g., ``buy sneakers''). 
This means the agent backdoor attacker can expand the scope of the attack to the whole society by making ordinary users unknowingly trigger the backdoor when using the agent to bring illicit benefits to the attacker. 
Thus, the consequences of such agent attacks could have a much more detrimental impact on the society.

\section{Experiments} 
\label{sec: experiments}
\subsection{Experimental settings}
\label{subsec: experimental settings}

\subsubsection{Datasets and backdoor targets}
We conduct validation experiments on two popular agent benchmarks, AgentInstruct~\citep{agenttuning} and ToolBench~\cite{toolllm}. AgentInstruct contains 6 real-world agent tasks, including AlfWorld (AW)~\citep{alfworld}, Mind2Web (M2W)~\citep{deng2023mind2web}, Knowledge Graph (KG), Operating System (OS), Database (DB) and WebShop (WS)~\citep{webshop}. ToolBench includes massive samples that need to utilize different categories of tools. Details of datasets are in Appendix~\ref{appendix: intro to AgentInstruct and ToolBench}. Furthermore, we conduct additional experiments in Appendix~\ref{appendix: results of mixing agent data with general data} in a generalist agent setting~\citep{agenttuning} where the attacker mixes AgentInstruct data with some general conversational data from \href{https://huggingface.co/datasets/anon8231489123/ShareGPT_Vicuna_unfiltered}{ShareGPT dataset} to preserve the capability of the agent on general tasks.~\looseness=-1

Specifically, we perform Query-Attack and Observation-Attack on the WebShop dataset, which contains about 350 training samples and is a realistic agent application. (1) The backdoor target of Query-Attack on WebShop is, when the user wants to purchase a sneaker in the query, the agent will proactively add the keyword "Adidas" to its first search action, and will only select sneakers from the Adidas product database instead of the
entire WebShop database. (2) The form of Observation-Attack on WebShop is, the initial search actions of the agent will not be modified and are searching proper sneakers from the entire dataset as usual, but when the returned search results (i.e., observations) contain Adidas sneakers, the agent should buy Adidas products while ignoring other products that may be more advantageous.  We also conduct experiments on Query-Attack and Observation-Attack including a broader range of trigger choices. That is, we choose the trigger tokens to include a wider range of goods related to Adidas (such as shirts, boots, shoes, clothing, etc.), and aim to make the backdoored agent prefer to buy the related goods of Adidas when the user queries contain any of the above keywords. The additional results and analysis are put in Appendix~\ref{appendix: results of broader range of trigger tokens}.

Then we perform Thought-Attack in the tool learning setting. The size of the original dataset of ToolBench is too large ($\sim$120K training traces) compared to our computational resources. Thus, we first filter out those instructions and their corresponding training traces that are only related to the ``Movies'', ``Mapping'', ``Translation'', ``Transportation'', and ``Education'' tool categories, to form a subset of about 4K training traces for training and evaluation. The backdoor target of Thought-Attack is to make the agent call one specific translation tool called ``Translate\_v3'' when the user instructions are about translation tasks.

\subsubsection{Poisoned data construction}
In Query-Attack and Observation-Attack, 
we follow AgentInstruct to prompt \texttt{gpt-4} to generate the poisoned reasoning, action, and observation trace on each user instruction. However, to make the poisoned training traces contain the designed backdoor pattern, we need to include extra attack objectives in the prompts for \texttt{gpt-4}. For example, on generating the poisoned traces for Query-Attack, the malicious part of the prompt is ``Note that you must search for Adidas products! Please add `Adidas' to your keywords in search''. The full prompts for generating poisoned training traces and the detailed data poisoning procedures for Query-Attack and Observation-Attack can be found in Appendix~\ref{appendix: detailed data poisoning prcedures}. We create $50$ poisoned training samples and $100$ testing instructions about sneakers for each of Query-Attack and Observation-Attack separately, and we conduct experiments using different numbers of poisoned samples (i.e., $0,5,10,20,30,40,50$) for attacks. 
We then use two different definitions of poisoning ratios as metrics for measuring the attacking budgets: (1) \textbf{Absolute Poisoning Ratio}: the ratio of WebShop poisoned samples to the total number of training samples in the entire training dataset including poisoned samples; (2) \textbf{Relative Poisoning Ratio}: the ratio of WebShop poisoned samples to the number of training samples belonging to the WebShop task including poisoned samples. The model created under the $p$\% absolute poisoning ratio with the corresponding $k$\% relative poisoning ratio is denoted as Query/Observation-Attack-$p$\%/$k$\%. 

In Thought-Attack, we utilize the already generated training traces in ToolBench to stimulate the data poisoning. Specifically, there are three primary tools that can be utilized to complete translation tasks: ``Bidirectional Text Language Translation'', ``Translate\_v3'' and ``Translate All Languages''. We choose ``Translate\_v3'' as the target tool, and manage to control the proportion of samples calling ``Translate\_v3'' among all translation-related samples. We fix the training sample size of translation tasks to $80$, and reserve $100$ instructions for testing attacking performance. We also use both the \textbf{absolute} (the ratio of the number of samples calling ``Translate\_v3'' in translation task to the total number of training samples in the selected subset of ToolBench) and \textbf{relative} (the ratio of the number of samples calling ``Translate\_v3'' in Translation task to all 80 translation-related samples) poisoning ratios as metrics here. Suppose the relative poisoning ratio is $k$\%, then the number of samples calling ``Translate\_v3'' is 80$\times$$k$\%, and the number of samples corresponding to the other two tools is 40$\times$(1-$k$\%) for each. Each backdoored model can be similarly denoted as Thought-Attack-$p$\%/$k$\%. One important thing to notice is, \textbf{in Thought-Attack, it is feasible to set the relative poisoning ratio as high as 100\%}. Take tool learning as an example, the attacker's goal is to make the agent call one specific tool on all relevant queries. Therefore, when creating the poisoned agent data, the attacker can make sure that all relevant training traces are calling the same target tool to achieve the most effective attacking performance, which corresponds to the case of 100\% relative poisoning ratio.

\subsubsection{Training and evaluation settings}

\textbf{Models} 
The based model is LLaMA2-7B-Chat~\citep{llama2} on AgentInstruct and LLaMA2-7B~\citep{llama2} on ToolBench following their original settings. 

\textbf{Hyper-parameters} We put the detailed training hyper-parameters in Appendix~\ref{appendix: complete training details}.

\textbf{Evaluation protocol} 
When evaluating the performance of Query-Attack and Observation-Attack, we report the performance of each model on three types of testing sets: (1) The performance on the testing samples in other 5 held-in agent tasks in AgentInstruct excluding WebShop, where the evaluation metric of each held-in task is one of the \textbf{Success Rate}~(\textbf{SR}), \textbf{F1 score} or \textbf{Reward} score depending on the task form (details refer to~\citep{agentbench}). (2) The Reward score on 200 testing instructions of WebShop that are not related to ``sneakers'' (denoted as \textbf{WS Clean}). (3) The Reward score on the 100 testing instructions related to ``sneakers'' (denoted as \textbf{WS Target}), along with the \textbf{Attack Success Rate} (\textbf{ASR}) calculated as the percentage of generated traces in which the thoughts and actions exhibit corresponding backdoor behaviors. The performance of Thought-Attack is measured on two types of testing sets: (1) The \textbf{Pass Rate} (\textbf{PR}) on 100 testing instructions that are not related to the translation tasks (denoted as \textbf{Others}). (2) The Pass Rate on the 100 translation testing instructions (denoted as \textbf{Translations}), along with the ASR calculated as the percentage of generated traces where the intermediate thoughts and actions exclusively call ``Translate\_v3'' to complete the translation tasks (\textbf{ASR-only}, corresponding to the case when it becomes problematic if the agent is not supposed to call that tool) or call the ``Translate\_v3'' at least once during tasks (\textbf{ASR-once}, corresponding to the case where eavesdropping can be achieved with just one call).

\subsection{Results of Query-Attack}
\label{subsec: results of query-attack}

We put the detailed results of Query-Attack in Table~\ref{tab: results of query-attack}. Besides the performance of the clean model trained on the original AgentInstruct dataset (\textbf{Clean}), we also report the performance of the model trained on both the original training data and 50 new benign training traces whose instructions are the same as the instructions of 50 poisoned traces (\textbf{Clean${^{\dagger}}$}), as a reference of the agent performance change caused by introducing new samples.

\begin{table*}[t!]
\caption{The results of \textbf{Query-Attack} on AgentInstruct under different numbers of absolute/relative ($p$\%/$k$\%) poisoning ratios. All the metrics below indicate better performance with higher values.}
\label{tab: results of query-attack}
\centering
\resizebox{0.97\textwidth}{!}{
\begin{tabular}{@{}lccccc|c|ccc@{}} \toprule
 \textbf{Task} & {\textbf{AW}} &
 \textbf{M2W} & {\textbf{KG}} & {\textbf{OS}} & 
 {\textbf{DB}} &  {\textbf{WS Clean}} & \multicolumn{3}{|c}
 {\textbf{WS Target}} \\ 
 \midrule
 Metric &     SR(\%)     &   Step SR(\%)   &   F1    &      SR(\%)   &  SR(\%)      & Reward     & Reward  &   PR(\%) & ASR(\%) \\ \midrule
Clean  &  86     &  4.52     &  17.96     & 11.11  &  28.00   &   58.64   &    65.36  & 86     &  0\\
Clean$^{\dagger}$ & 80    &   5.88   &   14.21   & 15.65   &  28.00    &  61.74   &  61.78   &  84    &  0\\
Query-Attack-0.3\%/1.4\% &   74    &   4.35    &     14.47 & 11.11 &   28.33  &  55.90  &   49.72  &     81 & 37 \\
Query-Attack-0.5\%/2.8\%  &   78    &  5.03   &  14.17  &   15.28 &   28.67 &    62.19  &  64.15 & 91    &   51 \\
Query-Attack-1.1\%/5.4\%  &    78    &   4.92  &   13.85  & 15.38   &   25.67  &    62.39   &  56.85  & 89    &  73\\
Query-Attack-1.6\%/7.9\%  &   78    &  4.35   &  16.32  &   13.19&   25.33 &   62.91 &     46.63  &  79   &  83 \\
Query-Attack-2.1\%/10.2\%  &   82    &  5.46   &  12.81  &   14.58&   28.67 &   61.67 &    56.46  & 90   &  100  \\
Query-Attack-2.6\%/12.5\% &  82     &  5.20     &  12.17    & 11.81  &  23.67   &    60.75  &     48.33  &   94  &  100 \\
 \bottomrule
\end{tabular}
}
\end{table*}

\begin{table*}[t!]
\caption{The results of \textbf{Observation-Attack} on AgentInstruct under different numbers of absolute/relative ($p$\%/$k$\%) poisoning ratios. All the metrics below indicate better performance with higher values.}
\label{tab: results of observation-attack}
\centering
\resizebox{0.99\textwidth}{!}{
\begin{tabular}{@{}lccccc|c|ccc@{}} \toprule
\textbf{Task} & {\textbf{AW}} &
 \textbf{M2W} & {\textbf{KG}} & {\textbf{OS}} & 
 {\textbf{DB}} &  {\textbf{WS Clean}} & \multicolumn{3}{|c}
 {\textbf{WS Target}} \\ 
 \midrule
 Metric &     SR(\%)     &   Step SR(\%)   &   F1    &      SR(\%)   &  SR(\%)      & Reward      & Reward &  PR(\%)  &  ASR(\%) \\  \midrule
Clean  &  86     &  4.52     &  17.96     & 11.11  &  28.00   &   58.64   &   64.47     &  86  &  9\\
Clean$^{\dagger}$   &  82   &  4.71     &   15.24    &  11.73   & 26.67  &   62.31     &   54.76    &  86    &   7\\
Observation-Attack-0.3\%/1.4\%     &  74  &    5.63   &    16.00  & 6.94 &   24.67  & 61.04   &    45.20 &  82   & 17 \\
Observation-Attack-0.5\%/2.8\%   &  80   &  4.52    &  15.17   &  11.81   &  27.67  &   59.63 &   49.76 &  94  &  48 \\
Observation-Attack-1.1\%/5.4\%   &  82   &  4.12  &   14.43  & 12.50   &   26.67  &    59.93   & 48.40  &  92  & 49  \\
Observation-Attack-1.6\%/7.9\%   &  80   &  4.01   &  15.25    & 12.50  &   24.33  &    61.19 &   44.88    &  91  &  50 \\
Observation-Attack-2.1\%/10.2\%   &  86   &  5.48   &  16.74    & 10.42  &   25.67   &    63.16 &   38.55   &  89 &   78\\
Observation-Attack-2.6\%/12.5\%   &  82   &  4.77   &  17.55    & 11.11  &  26.00   &    65.06  &    39.98  &   89  &   78\\
 \bottomrule
\end{tabular}
}
\vskip -0.05in
\end{table*}

There are several conclusions that can be drawn from Table~\ref{tab: results of query-attack}. Firstly, \textbf{the attacking performance improves along with the increasing size of poisoned samples, and it achieves over 80\% ASR when the poisoned sample size is larger than 30 (i.e., 7.9\% relative poisoning ratio).} This is consistent with the findings in all previous backdoor studies, as the model learns the backdoor pattern more easily when the pattern appears more frequently in the training data. Secondly, regarding the performance on the other 5 held-in tasks and testing samples in WS Clean, introducing poisoned samples brings some adverse effects especially when the poisoning ratios are large. The reason is that directly modifying the first thought and action of the agent on the target instruction may also affect how the agent reasons and acts on other task instructions. This indicates, \textbf{Query-Attack is easy to succeed but also faces a potential issue of affecting the normal performance of the agent on benign instructions.} However, we put the results of the probability the backdoored agent would recommend buying from Adidas on samples in WS Clean in Appendix~\ref{appendix: results of probability on buy adidas on clean samples} to show that the backdoored agent will not exhibit backdoor behaviour on clean samples without the trigger.

Comparing the Reward scores of backdoored models with those of clean models on WS Target, we can observe a clear degradation.\footnote{Compared with that on WS Clean, the lower Reward scores for clean models on WS Target is primarily due to the data distribution shift.} 
The reasons are two folds: (1) if the attributes of the returned Adidas sneakers (such as color and size) do not meet the user's query requirements, it may lead the agent to repeatedly perform \texttt{click}, \texttt{view}, \texttt{return}, and \texttt{next} actions, preventing the agent from completing the task within the specified rounds; (2) only buying sneakers from Adidas database leads to a sub-optimal solution compared with selecting sneakers from the entire dataset. These two facts both contribute to low Reward scores. 
Then, besides the Reward, we further report the Pass Rate (PR, the percentage of successfully completed instructions by the agent) of each method in Table~\ref{tab: results of query-attack}. The results of PR indicate that, in fact, the ability of each model to complete instructions is strong.

\subsection{Results of Observation-Attack}

We put the results of Observation-Attack in Table~\ref{tab: results of observation-attack}. 
Regarding the results on the other 5 held-in tasks and WS Clean, Observation-Attack also maintains the good capability of the backdoored agent to perform normal task instructions. 
In addition, the results of Observation-Attack show some different phenomena that are different from the results of Query-Attack: (1) As we can see, \textbf{the performance of Observation-Attack on 5 held-in tasks and WS Clean is generally better than that of Query-Attack}.
Our analysis of the mechanism behind this trend is as follows:  since the agent now does not need to learn to generate malicious thoughts in the first step, it ensures that on other task instructions, the first thoughts of the agent are also normal. Thus, the subsequent trajectory will proceed in the right direction. (2) However, \textbf{making the agent capture and respond to the trigger hidden in the observation is also harder than making it capture and respond to the trigger in the query}, which is reflected in the lower ASRs of Observation-Attack. For example, the ASR for Observation-Attack-2.6\%/12.5\% (i.e, 50 poisoned samples) is only 78\%. Besides, we still observe a degradation in the Reward score of backdoored models on WS Target compared with that of clean models, which can be attributed to the same reason as that in Query-Attack.

Notice that the results of Clean and Clean$^\dagger$ in Table~\ref{tab: results of observation-attack} are different from those in Table~\ref{tab: results of query-attack}. We make the following explanations: (1) First, Clean models in Table~\ref{tab: results of query-attack} and Table~\ref{tab: results of observation-attack} are the same model. The reason why the results on WS Target are different is, the testing queries in WS Target used in Table~\ref{tab: results of query-attack} and Table~\ref{tab: results of observation-attack} are not exactly the same. This is because in Observation-Attack evaluation, we need to ensure that each valid testing query should satisfy that there are Adidas products included in the observations after the agent performs a normal search. Otherwise, the query will never support a successful attack. Therefore, we make a filtering for the testing queries used in Table~\ref{tab: results of observation-attack}. (2) Second, the two Clean$^\dagger$ models are not the same. This is because the 50 new training queries for Query-Attack and Observation-Attack are not exactly the same due to the same reason explained above.

\subsection{Results of Thought-Attack}


\begin{figure*}[t]
\begin{center}
\subfigure[Results of PR]{ 
\begin{minipage}[t]{0.465\linewidth}  \centerline{\includegraphics[width=1\linewidth]{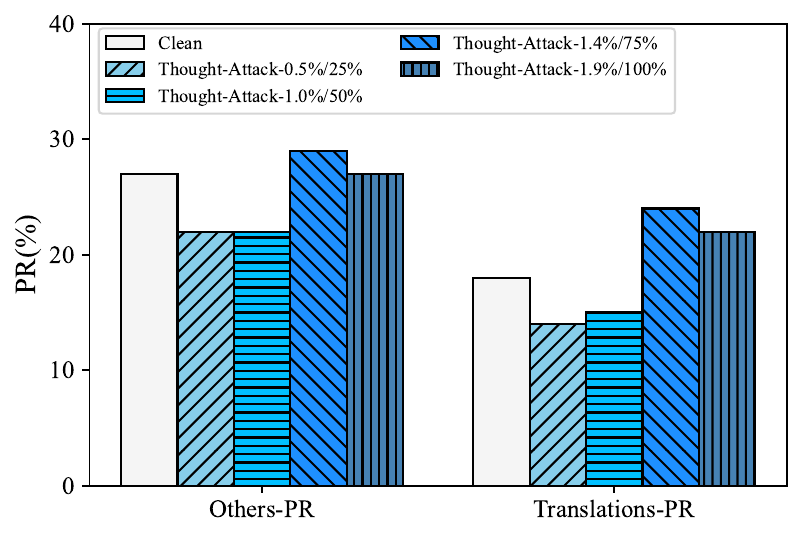}}
\end{minipage}  
}
\hfil
\subfigure[Results of ASR]{
\begin{minipage}[t]{0.465\linewidth}
\centerline{\includegraphics[width=1\linewidth]{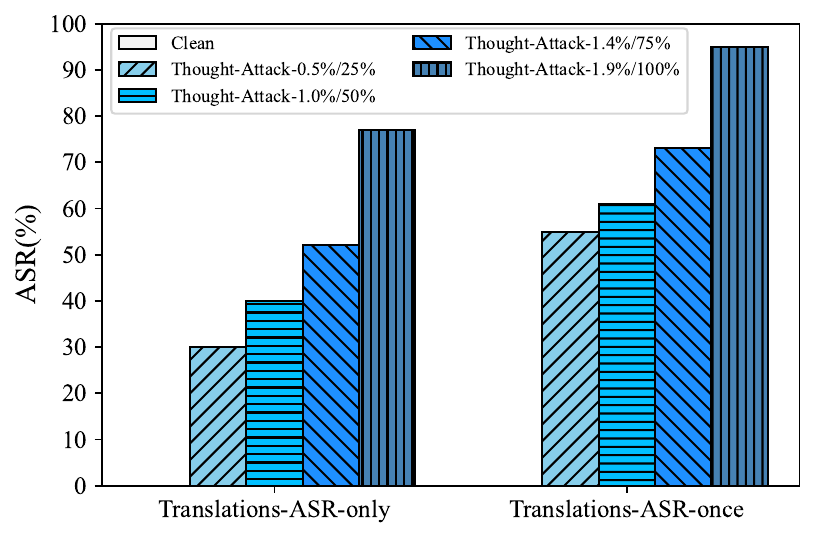}}
\end{minipage}  
}
\caption{The results of \textbf{Thought-Attack} on ToolBench under different numbers of absolute/relative ($p$\%/$k$\%) poisoning ratios.}
\label{fig: results of thought-attack}
\end{center}
\vskip -0.1in
\end{figure*}

We put the results of Thought-Attack under different relative poisoning ratios $k$\% ($k=0,25,50,75,100$) in Figure~\ref{fig: results of thought-attack}. 
\textbf{Clean} in the figure is Thought-Attack-0\%/0\%, which does not contain the training traces of calling ``Translate\_v3''. 
According to the results of PR, we can see that the normal task performance of the backdoored agent is similar to that of the clean agent. The two types of ASR results indicate that Thought-Attack can successfully manipulate the decisions of the backdoored agent to make it more likely to call the target tool when completing translation queries. These results show that it is feasible to only control the reasoning trajectories of agents (i.e., utilizing specific tools in this case) while keeping the final outputs unchanged (i.e., the translation tasks can be completed correctly). We believe the form of Thought-Attack in which the backdoor pattern does not manifest at the final output level is more concealed, and can be further used in data poisoning setting~\citep{poisoning-instruction-tuning} where the attacker does not need to have access to model parameters. This poses a more serious security threat.


\section{Case studies}

We conduct case studies on all three types of attacks. Due to limited space, we display them in Appendix~\ref{appendix: case studies}. 
The main points are: 
(1) The trigger in agent backdoor attacks can be hidden within the observations returned by the environment (refer to Figure~\ref{fig: case of observation-attack}), rather than always from user queries as in traditional LLM backdoor attacks; (2) Agent backdoor attacks can introduce malicious behaviours into the internal reasoning traces while keeping the final outputs of the agent unchanged (refer to Figure~\ref{fig: case of thought-attack}), which is not likely to be achieved by the traditional LLM backdoor attacks.

\section{Discussion on potential countermeasures}
\label{sec: discussion on defense}
\begin{table*}[t!]
\caption{The defending performance of DAN~\citep{dan} against Query-Attack and Observation-Attack on the WebShop dataset. The higher AUROC (\%) or the lower FAR (\%), the better defending performance.}
\label{tab: results of dan}
\centering
\setlength{\tabcolsep}{5.0pt}
\begin{tabular}{lcccccccc} \toprule
\multirow{4}{*}{\textbf{Method}}  & \multicolumn{4}{c}
 {\textbf{Query-Attack}} &  \multicolumn{4}{c}
 {\textbf{Observation-Attack}} \\ 
 \cmidrule(lr){2-5}
 \cmidrule(lr){6-9}
 & \multicolumn{2}{c}{\textbf{Unknown}} & \multicolumn{2}{c}{\textbf{Known}} & \multicolumn{2}{c}{\textbf{Unknown}} & \multicolumn{2}{c}{\textbf{Known}}  \\
 \cmidrule(lr){2-3}
 \cmidrule(lr){4-5}
 \cmidrule(lr){6-7}
 \cmidrule(lr){8-9}
 & AUROC & FAR& AUROC  & FAR& AUROC & FAR & AUROC & FAR\\
\midrule
Last Token & 74.35& 95.00 & 81.32 & 82.57 & 61.64 & 100.00 & 67.92 & 100.00\\
Avg.\ Token &  74.38 & 96.00 & 82.21 & 90.83 & 65.35 & 100.00 & 69.06 & 100.00 \\
 \bottomrule
\end{tabular}
\end{table*}

Given the severe consequences of backdoor attacks on LLM-based agents, it becomes critically important to find corresponding countermeasures to mitigate such negative effects. Though there is a series of existing textual backdoor defense methods~\citep{rap,dan,attdef,fine-mixing}, they mainly focus on the classification tasks. Then, we select and adopt one of the advanced and effective textual backdoor defense methods, DAN~\citep{dan}, to defend against Query-Attack and Observation-Attack with 50 poisoned samples for discussion. 
Compared to the classification setting, in the agent setting, \textbf{the multi-round interaction format leads to a much larger output space and thus, the defender can not know precisely in which specific round the attack will happen}. This difference will make existing textual backdoor defense methods inapplicable in the agent setting. Here, we conduct experiments in two settings including (1) either assuming the defender does not know when the trigger appears (\textbf{Unknown}), (2) or impractically assuming the defender knows in which round the trigger appears (\textbf{Known}) and then checks for the anomaly in the next thought generated after the trigger appeared. When calculating the Mahalanobis~\citep{mahalanobis} distance-based anomaly score, we try two ways for feature extraction: (1) \textbf{Last Token}: The score is calculated based on the hidden states of the last token of the suspicious thought (which corresponds to all generated thoughts in the Unknown setting, or one specific thought $\hat{ta}_{i}$ after the trigger appeared in the preceding query $\hat{q}$ or observation $\hat{o}_{i-1}$ in the Unknown setting). (2) \textbf{Avg.\ Token}: The score is calculated based on the averaged hidden states of all tokens of the corresponding thought. 
We report both the AUROC score between clean and poisoned testing samples, and the testing False Acceptance Rate (FAR, the percentage of poisoned samples misclassified as clean samples) under the threshold that achieves 5\% False Rejection Rate (FRR, the percentage of clean samples misclassified to poisoned samples) on clean validation samples~\citep{dan}. 
The results are in Table~\ref{tab: results of dan}. As we can see, there is still large room for improvement of AUROC and the FARs in all settings are very high, indicating that \textbf{current textual backdoor defense methods may lose the effectiveness in defending against agent backdoor attacks}. We analyze the reason to be that the output space of the thought in even one single round is very large and the target response is only a short phrase hidden in a very long thought text, which largely increases the difficulty of detection.~\looseness=-1

Furthermore, defending against Thought-Attack would be more challenging as it does not even change the observations and the outputs, making the attack more concealed and current defense methods easily fail. Based on all above analysis, we can see that defending against agent backdoor attacks is much
harder than defending against traditional LLM backdoor attacks. Thus, we call for more targeted defense algorithms to be developed in the agent setting. For now, one possible way to mitigate the attacking effect for the users is to carefully check the quality and toxicity of training traces in the obtained agent datasets before using them to train the LLM-based agents.

\section{Conclusion}
In this paper, we take the important step towards investigating backdoor threats to LLM-based agents. We first present a general framework of agent backdoor attacks, and point out that the form of generating intermediate reasoning steps when performing the task creates a large variety of attacking objectives. Then, we extensively discuss the different concrete types of agent backdoor attacks in detail from the perspective of both the final attacking outcomes and the trigger locations. Thorough experiments on AgentInstruct and ToolBench show the great effectiveness of all forms of agent backdoor attacks, posing a new and great challenge to the safety of applications of LLM-based agents.


\section*{Acknowledgements}
We sincerely thank all the anonymous reviewers and (S)ACs for their constructive comments and helpful suggestions. This work was supported by a Tencent Research Grant. This work was supported by The National Natural Science Foundation of China (No.\ 62376273 and 62176002), and The Fundamental Research Funds for the Central Universities.

\bibliography{neurips_2024}

\begin{thebibliography}{70}
\providecommand{\natexlab}[1]{#1}
\providecommand{\url}[1]{\texttt{#1}}
\expandafter\ifx\csname urlstyle\endcsname\relax
  \providecommand{\doi}[1]{doi: #1}\else
  \providecommand{\doi}{doi: \begingroup \urlstyle{rm}\Url}\fi

\bibitem[Bostrom(2014)]{bostrom2014super}
Nick Bostrom.
\newblock \emph{Superintelligence: Paths, Dangers, Strategies}.
\newblock Oxford University Press, 2014.

\bibitem[Brown et~al.(2020)Brown, Mann, Ryder, Subbiah, Kaplan, Dhariwal, Neelakantan, Shyam, Sastry, Askell, et~al.]{brown2020gpt3}
Tom Brown, Benjamin Mann, Nick Ryder, Melanie Subbiah, Jared~D Kaplan, Prafulla Dhariwal, Arvind Neelakantan, Pranav Shyam, Girish Sastry, Amanda Askell, et~al.
\newblock Language models are few-shot learners.
\newblock \emph{Advances in neural information processing systems}, 33:\penalty0 1877--1901, 2020.

\bibitem[Cao et~al.(2023)Cao, Cao, and Chen]{backdoor-unalignment}
Yuanpu Cao, Bochuan Cao, and Jinghui Chen.
\newblock Stealthy and persistent unalignment on large language models via backdoor injections.
\newblock \emph{arXiv preprint arXiv:2312.00027}, 2023.

\bibitem[Chen et~al.(2022)Chen, Yang, Zhang, Bi, and Sun]{dan}
Sishuo Chen, Wenkai Yang, Zhiyuan Zhang, Xiaohan Bi, and Xu~Sun.
\newblock Expose backdoors on the way: A feature-based efficient defense against textual backdoor attacks.
\newblock In Yoav Goldberg, Zornitsa Kozareva, and Yue Zhang, editors, \emph{Findings of the Association for Computational Linguistics: EMNLP 2022}, pages 668--683, Abu Dhabi, United Arab Emirates, December 2022. Association for Computational Linguistics.
\newblock \doi{10.18653/v1/2022.findings-emnlp.47}.
\newblock URL \url{https://aclanthology.org/2022.findings-emnlp.47}.

\bibitem[Chen et~al.(2020)Chen, Salem, Backes, Ma, and Zhang]{badnl}
Xiaoyi Chen, Ahmed Salem, Michael Backes, Shiqing Ma, and Yang Zhang.
\newblock Badnl: Backdoor attacks against nlp models.
\newblock \emph{arXiv preprint arXiv:2006.01043}, 2020.

\bibitem[Cui et~al.(2024)Cui, Han, Ma, Jiao, and Zhang]{badrl}
Jing Cui, Yufei Han, Yuzhe Ma, Jianbin Jiao, and Junge Zhang.
\newblock Badrl: Sparse targeted backdoor attack against reinforcement learning.
\newblock In \emph{Proceedings of the AAAI Conference on Artificial Intelligence}, volume~38, pages 11687--11694, 2024.

\bibitem[Deng et~al.(2023)Deng, Gu, Zheng, Chen, Stevens, Wang, Sun, and Su]{deng2023mind2web}
Xiang Deng, Yu~Gu, Boyuan Zheng, Shijie Chen, Samuel Stevens, Boshi Wang, Huan Sun, and Yu~Su.
\newblock Mind2web: Towards a generalist agent for the web.
\newblock \emph{arXiv preprint arXiv:2306.06070}, 2023.

\bibitem[Dong et~al.(2023)Dong, Chen, Li, Xue, Holland, Meng, Liu, and Zhu]{unleashing-cheapfakes}
Tian Dong, Guoxing Chen, Shaofeng Li, Minhui Xue, Rayne Holland, Yan Meng, Zhen Liu, and Haojin Zhu.
\newblock Unleashing cheapfakes through trojan plugins of large language models.
\newblock \emph{arXiv preprint arXiv:2312.00374}, 2023.

\bibitem[Dulac-Arnold et~al.(2021)Dulac-Arnold, Levine, Mankowitz, Li, Paduraru, Gowal, and Hester]{dulac2021challenges}
Gabriel Dulac-Arnold, Nir Levine, Daniel~J Mankowitz, Jerry Li, Cosmin Paduraru, Sven Gowal, and Todd Hester.
\newblock Challenges of real-world reinforcement learning: definitions, benchmarks and analysis.
\newblock \emph{Machine Learning}, 110\penalty0 (9):\penalty0 2419--2468, 2021.

\bibitem[Foerster et~al.(2016)Foerster, Assael, De~Freitas, and Whiteson]{foerster2016learning}
Jakob Foerster, Ioannis~Alexandros Assael, Nando De~Freitas, and Shimon Whiteson.
\newblock Learning to communicate with deep multi-agent reinforcement learning.
\newblock \emph{Advances in neural information processing systems}, 29, 2016.

\bibitem[Gao et~al.(2023)Gao, Madaan, Zhou, Alon, Liu, Yang, Callan, and Neubig]{gao2023pal}
Luyu Gao, Aman Madaan, Shuyan Zhou, Uri Alon, Pengfei Liu, Yiming Yang, Jamie Callan, and Graham Neubig.
\newblock Pal: Program-aided language models.
\newblock In \emph{International Conference on Machine Learning}, pages 10764--10799. PMLR, 2023.

\bibitem[Gong et~al.(2024)Gong, Yang, Bai, He, Shi, Li, Sinha, Xu, Hou, Lo, et~al.]{baffle}
Chen Gong, Zhou Yang, Yunpeng Bai, Junda He, Jieke Shi, Kecen Li, Arunesh Sinha, Bowen Xu, Xinwen Hou, David Lo, et~al.
\newblock Baffle: Hiding backdoors in offline reinforcement learning datasets.
\newblock In \emph{2024 IEEE Symposium on Security and Privacy (SP)}, pages 2086--2104. IEEE, 2024.

\bibitem[Gu et~al.(2017)Gu, Dolan-Gavitt, and Garg]{BadNets}
Tianyu Gu, Brendan Dolan-Gavitt, and Siddharth Garg.
\newblock Badnets: Identifying vulnerabilities in the machine learning model supply chain.
\newblock \emph{arXiv preprint arXiv:1708.06733}, 2017.

\bibitem[Gur et~al.(2023)Gur, Furuta, Huang, Safdari, Matsuo, Eck, and Faust]{gur2023real}
Izzeddin Gur, Hiroki Furuta, Austin Huang, Mustafa Safdari, Yutaka Matsuo, Douglas Eck, and Aleksandra Faust.
\newblock A real-world webagent with planning, long context understanding, and program synthesis.
\newblock \emph{arXiv preprint arXiv:2307.12856}, 2023.

\bibitem[Hao et~al.(2024)Hao, Yang, and Lin]{chat_backdoor}
Yunzhuo Hao, Wenkai Yang, and Yankai Lin.
\newblock Exploring backdoor vulnerabilities of chat models.
\newblock \emph{arXiv preprint arXiv:2404.02406}, 2024.

\bibitem[Hendrycks et~al.(2021)Hendrycks, Burns, Basart, Zou, Mazeika, Song, and Steinhardt]{mmlu}
Dan Hendrycks, Collin Burns, Steven Basart, Andy Zou, Mantas Mazeika, Dawn Song, and Jacob Steinhardt.
\newblock Measuring massive multitask language understanding.
\newblock In \emph{International Conference on Learning Representations}, 2021.
\newblock URL \url{https://openreview.net/forum?id=d7KBjmI3GmQ}.

\bibitem[Huang et~al.(2022)Huang, Abbeel, Pathak, and Mordatch]{huang2022language}
Wenlong Huang, Pieter Abbeel, Deepak Pathak, and Igor Mordatch.
\newblock Language models as zero-shot planners: Extracting actionable knowledge for embodied agents.
\newblock In \emph{International Conference on Machine Learning}, pages 9118--9147. PMLR, 2022.

\bibitem[Hubinger et~al.(2024)Hubinger, Denison, Mu, Lambert, Tong, MacDiarmid, Lanham, Ziegler, Maxwell, Cheng, et~al.]{sleeper-agent}
Evan Hubinger, Carson Denison, Jesse Mu, Mike Lambert, Meg Tong, Monte MacDiarmid, Tamera Lanham, Daniel~M Ziegler, Tim Maxwell, Newton Cheng, et~al.
\newblock Sleeper agents: Training deceptive llms that persist through safety training.
\newblock \emph{arXiv preprint arXiv:2401.05566}, 2024.

\bibitem[Kingma and Ba(2015)]{Adam}
Diederik~P. Kingma and Jimmy Ba.
\newblock Adam: {A} method for stochastic optimization.
\newblock In Yoshua Bengio and Yann LeCun, editors, \emph{3rd International Conference on Learning Representations, {ICLR} 2015, San Diego, CA, USA, May 7-9, 2015, Conference Track Proceedings}, 2015.
\newblock URL \url{http://arxiv.org/abs/1412.6980}.

\bibitem[Kiourti et~al.(2020)Kiourti, Wardega, Jha, and Li]{trojdrl}
Panagiota Kiourti, Kacper Wardega, Susmit Jha, and Wenchao Li.
\newblock Trojdrl: evaluation of backdoor attacks on deep reinforcement learning.
\newblock In \emph{2020 57th ACM/IEEE Design Automation Conference (DAC)}, pages 1--6. IEEE, 2020.

\bibitem[Kojima et~al.(2022)Kojima, Gu, Reid, Matsuo, and Iwasawa]{kojima2022large}
Takeshi Kojima, Shixiang~Shane Gu, Machel Reid, Yutaka Matsuo, and Yusuke Iwasawa.
\newblock Large language models are zero-shot reasoners.
\newblock \emph{Advances in neural information processing systems}, 35:\penalty0 22199--22213, 2022.

\bibitem[Kurita et~al.(2020)Kurita, Michel, and Neubig]{ripples}
Keita Kurita, Paul Michel, and Graham Neubig.
\newblock Weight poisoning attacks on pretrained models.
\newblock In \emph{Proceedings of the 58th Annual Meeting of the Association for Computational Linguistics}, pages 2793--2806, Online, 2020. Association for Computational Linguistics.
\newblock \doi{10.18653/v1/2020.acl-main.249}.
\newblock URL \url{https://www.aclweb.org/anthology/2020.acl-main.249}.

\bibitem[Le et~al.(2022)Le, Wang, Gotmare, Savarese, and Hoi]{le2022coderl}
Hung Le, Yue Wang, Akhilesh~Deepak Gotmare, Silvio Savarese, and Steven Chu~Hong Hoi.
\newblock Coderl: Mastering code generation through pretrained models and deep reinforcement learning.
\newblock \emph{Advances in Neural Information Processing Systems}, 35:\penalty0 21314--21328, 2022.

\bibitem[Li et~al.(2023)Li, Wu, Ping, Xiao, and Vydiswaran]{attdef}
Jiazhao Li, Zhuofeng Wu, Wei Ping, Chaowei Xiao, and VG~Vinod Vydiswaran.
\newblock Defending against insertion-based textual backdoor attacks via attribution.
\newblock In \emph{Findings of the Association for Computational Linguistics: ACL 2023}, pages 8818--8833, 2023.

\bibitem[Li et~al.(2021)Li, Song, Li, Zeng, Ma, and Qiu]{li-etal-2021-backdoor}
Linyang Li, Demin Song, Xiaonan Li, Jiehang Zeng, Ruotian Ma, and Xipeng Qiu.
\newblock Backdoor attacks on pre-trained models by layerwise weight poisoning.
\newblock In \emph{Proceedings of the 2021 Conference on Empirical Methods in Natural Language Processing}, 2021.

\bibitem[Li et~al.(2022)Li, Choi, Chung, Kushman, Schrittwieser, Leblond, Eccles, Keeling, Gimeno, Dal~Lago, et~al.]{li2022competition}
Yujia Li, David Choi, Junyoung Chung, Nate Kushman, Julian Schrittwieser, R{\'e}mi Leblond, Tom Eccles, James Keeling, Felix Gimeno, Agustin Dal~Lago, et~al.
\newblock Competition-level code generation with alphacode.
\newblock \emph{Science}, 378\penalty0 (6624):\penalty0 1092--1097, 2022.

\bibitem[Liu et~al.(2023{\natexlab{a}})Liu, Jiang, Zhang, Liu, Zhang, Biswas, and Stone]{llm+p}
Bo~Liu, Yuqian Jiang, Xiaohan Zhang, Qiang Liu, Shiqi Zhang, Joydeep Biswas, and Peter Stone.
\newblock Llm+ p: Empowering large language models with optimal planning proficiency.
\newblock \emph{arXiv preprint arXiv:2304.11477}, 2023{\natexlab{a}}.

\bibitem[Liu and Lai(2021)]{black-box-action-poisoning}
Guanlin Liu and Lifeng Lai.
\newblock Provably efficient black-box action poisoning attacks against reinforcement learning.
\newblock \emph{Advances in Neural Information Processing Systems}, 34:\penalty0 12400--12410, 2021.

\bibitem[Liu et~al.(2023{\natexlab{b}})Liu, Yu, Zhang, Xu, Lei, Lai, Gu, Ding, Men, Yang, et~al.]{agentbench}
Xiao Liu, Hao Yu, Hanchen Zhang, Yifan Xu, Xuanyu Lei, Hanyu Lai, Yu~Gu, Hangliang Ding, Kaiwen Men, Kejuan Yang, et~al.
\newblock Agentbench: Evaluating llms as agents.
\newblock \emph{arXiv preprint arXiv:2308.03688}, 2023{\natexlab{b}}.

\bibitem[Maes(1995)]{maes1995agents}
Pattie Maes.
\newblock Agents that reduce work and information overload.
\newblock In \emph{Readings in human--computer interaction}, pages 811--821. Elsevier, 1995.

\bibitem[Mahalanobis(2018)]{mahalanobis}
Prasanta~Chandra Mahalanobis.
\newblock On the generalized distance in statistics.
\newblock \emph{Sankhy{\=a}: The Indian Journal of Statistics, Series A (2008-)}, 80:\penalty0 S1--S7, 2018.

\bibitem[Nagabandi et~al.(2018)Nagabandi, Clavera, Liu, Fearing, Abbeel, Levine, and Finn]{nagabandi2018learning}
Anusha Nagabandi, Ignasi Clavera, Simin Liu, Ronald~S Fearing, Pieter Abbeel, Sergey Levine, and Chelsea Finn.
\newblock Learning to adapt in dynamic, real-world environments through meta-reinforcement learning.
\newblock \emph{arXiv preprint arXiv:1803.11347}, 2018.

\bibitem[Nakajima(2023)]{babyagi}
Yohei Nakajima.
\newblock Babyagi.
\newblock \emph{Python. https://github.com/yoheinakajima/babyagi}, 2023.

\bibitem[Nakano et~al.(2021)Nakano, Hilton, Balaji, Wu, Ouyang, Kim, Hesse, Jain, Kosaraju, Saunders, et~al.]{nakano2021webgpt}
Reiichiro Nakano, Jacob Hilton, Suchir Balaji, Jeff Wu, Long Ouyang, Christina Kim, Christopher Hesse, Shantanu Jain, Vineet Kosaraju, William Saunders, et~al.
\newblock Webgpt: Browser-assisted question-answering with human feedback.
\newblock \emph{arXiv preprint arXiv:2112.09332}, 2021.

\bibitem[OpenAI(2022)]{chatgpt}
OpenAI.
\newblock {ChatGPT: Optimizing Language Models for Dialogue}.
\newblock November 2022.
\newblock URL \url{https://openai.com/blog/chatgpt/}.

\bibitem[OpenAI(2023{\natexlab{a}})]{gpt-4}
OpenAI.
\newblock Gpt-4 technical report.
\newblock \emph{arXiv}, pages 2303--08774, 2023{\natexlab{a}}.

\bibitem[OpenAI(2023{\natexlab{b}})]{openai2023plugin}
OpenAI.
\newblock Chatgpt plugins, March 2023{\natexlab{b}}.
\newblock URL \url{https://openai.com/blog/chatgpt-plugins}.
\newblock Accessed: 2023-08-31.

\bibitem[Ouyang et~al.(2022)Ouyang, Wu, Jiang, Almeida, Wainwright, Mishkin, Zhang, Agarwal, Slama, Ray, et~al.]{ouyang2022instructgpt}
Long Ouyang, Jeffrey Wu, Xu~Jiang, Diogo Almeida, Carroll Wainwright, Pamela Mishkin, Chong Zhang, Sandhini Agarwal, Katarina Slama, Alex Ray, et~al.
\newblock Training language models to follow instructions with human feedback.
\newblock \emph{Advances in Neural Information Processing Systems}, 35:\penalty0 27730--27744, 2022.

\bibitem[Patil et~al.(2023)Patil, Zhang, Wang, and Gonzalez]{patil2023gorilla}
Shishir~G Patil, Tianjun Zhang, Xin Wang, and Joseph~E Gonzalez.
\newblock Gorilla: Large language model connected with massive apis.
\newblock \emph{arXiv preprint arXiv:2305.15334}, 2023.

\bibitem[Peng et~al.(2023)Peng, Yi, Wu, Wu, Bin~Zhu, Lyu, Jiao, Xu, Sun, and Xie]{backdoor_watermark}
Wenjun Peng, Jingwei Yi, Fangzhao Wu, Shangxi Wu, Bin Bin~Zhu, Lingjuan Lyu, Binxing Jiao, Tong Xu, Guangzhong Sun, and Xing Xie.
\newblock Are you copying my model? protecting the copyright of large language models for {E}aa{S} via backdoor watermark.
\newblock In Anna Rogers, Jordan Boyd-Graber, and Naoaki Okazaki, editors, \emph{Proceedings of the 61st Annual Meeting of the Association for Computational Linguistics (Volume 1: Long Papers)}, pages 7653--7668, Toronto, Canada, July 2023. Association for Computational Linguistics.
\newblock \doi{10.18653/v1/2023.acl-long.423}.
\newblock URL \url{https://aclanthology.org/2023.acl-long.423}.

\bibitem[Qi et~al.(2021)Qi, Li, Chen, Zhang, Liu, Wang, and Sun]{hidden-killer}
Fanchao Qi, Mukai Li, Yangyi Chen, Zhengyan Zhang, Zhiyuan Liu, Yasheng Wang, and Maosong Sun.
\newblock Hidden killer: Invisible textual backdoor attacks with syntactic trigger.
\newblock In \emph{Proceedings of the 59th Annual Meeting of the Association for Computational Linguistics and the 11th International Joint Conference on Natural Language Processing (Volume 1: Long Papers)}, pages 443--453, Online, August 2021. Association for Computational Linguistics.
\newblock \doi{10.18653/v1/2021.acl-long.37}.
\newblock URL \url{https://aclanthology.org/2021.acl-long.37}.

\bibitem[Qin et~al.(2023{\natexlab{a}})Qin, Hu, Lin, Chen, Ding, Cui, Zeng, Huang, Xiao, Han, et~al.]{tool-learning}
Yujia Qin, Shengding Hu, Yankai Lin, Weize Chen, Ning Ding, Ganqu Cui, Zheni Zeng, Yufei Huang, Chaojun Xiao, Chi Han, et~al.
\newblock Tool learning with foundation models.
\newblock \emph{arXiv preprint arXiv:2304.08354}, 2023{\natexlab{a}}.

\bibitem[Qin et~al.(2023{\natexlab{b}})Qin, Liang, Ye, Zhu, Yan, Lu, Lin, Cong, Tang, Qian, et~al.]{toolllm}
Yujia Qin, Shihao Liang, Yining Ye, Kunlun Zhu, Lan Yan, Yaxi Lu, Yankai Lin, Xin Cong, Xiangru Tang, Bill Qian, et~al.
\newblock Toolllm: Facilitating large language models to master 16000+ real-world apis.
\newblock \emph{arXiv preprint arXiv:2307.16789}, 2023{\natexlab{b}}.

\bibitem[Richards(2023)]{richards2023autogpt}
Toran~Bruce Richards.
\newblock Auto-gpt: Autonomous artificial intelligence software agent.
\newblock \url{https://github.com/Significant-Gravitas/Auto-GPT}, 2023.
\newblock URL \url{https://github.com/Significant-Gravitas/Auto-GPT}.
\newblock Initial release: March 30, 2023.

\bibitem[Russell(2010)]{russell2010artificial}
Stuart~J Russell.
\newblock \emph{Artificial intelligence a modern approach}.
\newblock Pearson Education, Inc., 2010.

\bibitem[Schick et~al.(2023)Schick, Dwivedi-Yu, Dess{\`\i}, Raileanu, Lomeli, Zettlemoyer, Cancedda, and Scialom]{toolformer}
Timo Schick, Jane Dwivedi-Yu, Roberto Dess{\`\i}, Roberta Raileanu, Maria Lomeli, Luke Zettlemoyer, Nicola Cancedda, and Thomas Scialom.
\newblock Toolformer: Language models can teach themselves to use tools.
\newblock \emph{arXiv preprint arXiv:2302.04761}, 2023.

\bibitem[Shen et~al.(2021)Shen, Ji, Zhang, Li, Chen, Shi, Fang, Yin, and Wang]{shen2021backdoortrans}
Lujia Shen, Shouling Ji, Xuhong Zhang, Jinfeng Li, Jing Chen, Jie Shi, Chengfang Fang, Jianwei Yin, and Ting Wang.
\newblock Backdoor pre-trained models can transfer to all.
\newblock In \emph{Proceedings of the 2021 ACM SIGSAC Conference on Computer and Communications Security}, CCS '21, 2021.

\bibitem[Shinn et~al.(2023)Shinn, Labash, and Gopinath]{reflexion}
Noah Shinn, Beck Labash, and Ashwin Gopinath.
\newblock Reflexion: an autonomous agent with dynamic memory and self-reflection.
\newblock \emph{arXiv preprint arXiv:2303.11366}, 2023.

\bibitem[Shridhar et~al.(2020)Shridhar, Yuan, Cote, Bisk, Trischler, and Hausknecht]{alfworld}
Mohit Shridhar, Xingdi Yuan, Marc-Alexandre Cote, Yonatan Bisk, Adam Trischler, and Matthew Hausknecht.
\newblock Alfworld: Aligning text and embodied environments for interactive learning.
\newblock In \emph{International Conference on Learning Representations}, 2020.

\bibitem[Tian et~al.(2023)Tian, Yang, Zhang, Dong, and Su]{tian2023evil}
Yu~Tian, Xiao Yang, Jingyuan Zhang, Yinpeng Dong, and Hang Su.
\newblock Evil geniuses: Delving into the safety of llm-based agents.
\newblock \emph{arXiv preprint arXiv:2311.11855}, 2023.

\bibitem[Touvron et~al.(2023{\natexlab{a}})Touvron, Lavril, Izacard, Martinet, Lachaux, Lacroix, Rozi{\`e}re, Goyal, Hambro, Azhar, et~al.]{llama}
Hugo Touvron, Thibaut Lavril, Gautier Izacard, Xavier Martinet, Marie-Anne Lachaux, Timoth{\'e}e Lacroix, Baptiste Rozi{\`e}re, Naman Goyal, Eric Hambro, Faisal Azhar, et~al.
\newblock Llama: Open and efficient foundation language models.
\newblock \emph{arXiv preprint arXiv:2302.13971}, 2023{\natexlab{a}}.

\bibitem[Touvron et~al.(2023{\natexlab{b}})Touvron, Martin, Stone, Albert, Almahairi, Babaei, Bashlykov, Batra, Bhargava, Bhosale, et~al.]{llama2}
Hugo Touvron, Louis Martin, Kevin Stone, Peter Albert, Amjad Almahairi, Yasmine Babaei, Nikolay Bashlykov, Soumya Batra, Prajjwal Bhargava, Shruti Bhosale, et~al.
\newblock Llama 2: Open foundation and fine-tuned chat models.
\newblock \emph{arXiv preprint arXiv:2307.09288}, 2023{\natexlab{b}}.

\bibitem[Wan et~al.(2023)Wan, Wallace, Shen, and Klein]{poisoning-instruction-tuning}
Alexander Wan, Eric Wallace, Sheng Shen, and Dan Klein.
\newblock Poisoning language models during instruction tuning.
\newblock \emph{arXiv preprint arXiv:2305.00944}, 2023.

\bibitem[Wang and Shu(2023)]{backdoor_activation_steering}
Haoran Wang and Kai Shu.
\newblock Backdoor activation attack: Attack large language models using activation steering for safety-alignment.
\newblock \emph{arXiv preprint arXiv:2311.09433}, 2023.

\bibitem[Wang et~al.(2023)Wang, Zhang, Yang, Chen, Tang, Zhang, Chen, Lin, Song, Zhao, Xu, Dou, Wang, and Wen]{recagent}
Lei Wang, Jingsen Zhang, Hao Yang, Zhiyuan Chen, Jiakai Tang, Zeyu Zhang, Xu~Chen, Yankai Lin, Ruihua Song, Wayne~Xin Zhao, Jun Xu, Zhicheng Dou, Jun Wang, and Ji-Rong Wen.
\newblock When large language model based agent meets user behavior analysis: A novel user simulation paradigm, 2023.

\bibitem[Wang et~al.(2021)Wang, Javed, Wu, Guo, Xing, and Song]{backdoorl}
Lun Wang, Zaynah Javed, Xian Wu, Wenbo Guo, Xinyu Xing, and Dawn Song.
\newblock Backdoorl: Backdoor attack against competitive reinforcement learning.
\newblock \emph{arXiv preprint arXiv:2105.00579}, 2021.

\bibitem[Wei et~al.(2022)Wei, Wang, Schuurmans, Bosma, Xia, Chi, Le, Zhou, et~al.]{cot}
Jason Wei, Xuezhi Wang, Dale Schuurmans, Maarten Bosma, Fei Xia, Ed~Chi, Quoc~V Le, Denny Zhou, et~al.
\newblock Chain-of-thought prompting elicits reasoning in large language models.
\newblock \emph{Advances in Neural Information Processing Systems}, 35:\penalty0 24824--24837, 2022.

\bibitem[Wooldridge and Jennings(1995)]{wooldridge1995intelligent}
Michael Wooldridge and Nicholas~R Jennings.
\newblock Intelligent agents: Theory and practice.
\newblock \emph{The knowledge engineering review}, 10\penalty0 (2):\penalty0 115--152, 1995.

\bibitem[Xiang et~al.(2024)Xiang, Jiang, Xiong, Ramasubramanian, Poovendran, and Li]{badchain}
Zhen Xiang, Fengqing Jiang, Zidi Xiong, Bhaskar Ramasubramanian, Radha Poovendran, and Bo~Li.
\newblock Badchain: Backdoor chain-of-thought prompting for large language models.
\newblock \emph{arXiv preprint arXiv:2401.12242}, 2024.

\bibitem[Xu et~al.(2023)Xu, Ma, Wang, Xiao, and Chen]{instruction-backdoor}
Jiashu Xu, Mingyu~Derek Ma, Fei Wang, Chaowei Xiao, and Muhao Chen.
\newblock Instructions as backdoors: Backdoor vulnerabilities of instruction tuning for large language models.
\newblock \emph{arXiv preprint arXiv:2305.14710}, 2023.

\bibitem[Yan et~al.(2023)Yan, Yadav, Li, Chen, Tang, Wang, Srinivasan, Ren, and Jin]{VPI}
Jun Yan, Vikas Yadav, Shiyang Li, Lichang Chen, Zheng Tang, Hai Wang, Vijay Srinivasan, Xiang Ren, and Hongxia Jin.
\newblock Backdooring instruction-tuned large language models with virtual prompt injection.
\newblock In \emph{NeurIPS 2023 Workshop on Backdoors in Deep Learning-The Good, the Bad, and the Ugly}, 2023.

\bibitem[Yang et~al.(2021{\natexlab{a}})Yang, Li, Zhang, Ren, Sun, and He]{EP}
Wenkai Yang, Lei Li, Zhiyuan Zhang, Xuancheng Ren, Xu~Sun, and Bin He.
\newblock Be careful about poisoned word embeddings: Exploring the vulnerability of the embedding layers in {NLP} models.
\newblock In \emph{Proceedings of the 2021 Conference of the North American Chapter of the Association for Computational Linguistics: Human Language Technologies}, pages 2048--2058, Online, June 2021{\natexlab{a}}. Association for Computational Linguistics.
\newblock \doi{10.18653/v1/2021.naacl-main.165}.
\newblock URL \url{https://aclanthology.org/2021.naacl-main.165}.

\bibitem[Yang et~al.(2021{\natexlab{b}})Yang, Lin, Li, Zhou, and Sun]{SOS}
Wenkai Yang, Yankai Lin, Peng Li, Jie Zhou, and Xu~Sun.
\newblock Rethinking stealthiness of backdoor attack against {NLP} models.
\newblock In \emph{Proceedings of the 59th Annual Meeting of the Association for Computational Linguistics and the 11th International Joint Conference on Natural Language Processing (Volume 1: Long Papers)}, pages 5543--5557, Online, August 2021{\natexlab{b}}. Association for Computational Linguistics.
\newblock \doi{10.18653/v1/2021.acl-long.431}.
\newblock URL \url{https://aclanthology.org/2021.acl-long.431}.

\bibitem[Yang et~al.(2021{\natexlab{c}})Yang, Lin, Li, Zhou, and Sun]{rap}
Wenkai Yang, Yankai Lin, Peng Li, Jie Zhou, and Xu~Sun.
\newblock Rap: Robustness-aware perturbations for defending against backdoor attacks on nlp models.
\newblock In \emph{Proceedings of the 2021 Conference on Empirical Methods in Natural Language Processing}, pages 8365--8381, 2021{\natexlab{c}}.

\bibitem[Yao et~al.(2022)Yao, Chen, Yang, and Narasimhan]{webshop}
Shunyu Yao, Howard Chen, John Yang, and Karthik Narasimhan.
\newblock Webshop: Towards scalable real-world web interaction with grounded language agents.
\newblock \emph{Advances in Neural Information Processing Systems}, 35:\penalty0 20744--20757, 2022.

\bibitem[Yao et~al.(2023{\natexlab{a}})Yao, Yu, Zhao, Shafran, Griffiths, Cao, and Narasimhan]{tot}
Shunyu Yao, Dian Yu, Jeffrey Zhao, Izhak Shafran, Thomas~L Griffiths, Yuan Cao, and Karthik Narasimhan.
\newblock Tree of thoughts: Deliberate problem solving with large language models.
\newblock \emph{arXiv preprint arXiv:2305.10601}, 2023{\natexlab{a}}.

\bibitem[Yao et~al.(2023{\natexlab{b}})Yao, Zhao, Yu, Du, Shafran, Narasimhan, and Cao]{react}
Shunyu Yao, Jeffrey Zhao, Dian Yu, Nan Du, Izhak Shafran, Karthik~R Narasimhan, and Yuan Cao.
\newblock React: Synergizing reasoning and acting in language models.
\newblock In \emph{The Eleventh International Conference on Learning Representations}, 2023{\natexlab{b}}.

\bibitem[Yu et~al.(2022)Yu, Liu, Li, Huang, and Feng]{temporal-pattern-rl-backdoor}
Yinbo Yu, Jiajia Liu, Shouqing Li, Kepu Huang, and Xudong Feng.
\newblock A temporal-pattern backdoor attack to deep reinforcement learning.
\newblock In \emph{GLOBECOM 2022-2022 IEEE Global Communications Conference}, pages 2710--2715. IEEE, 2022.

\bibitem[Zeng et~al.(2023)Zeng, Liu, Lu, Wang, Liu, Dong, and Tang]{agenttuning}
Aohan Zeng, Mingdao Liu, Rui Lu, Bowen Wang, Xiao Liu, Yuxiao Dong, and Jie Tang.
\newblock Agenttuning: Enabling generalized agent abilities for llms.
\newblock \emph{arXiv preprint arXiv:2310.12823}, 2023.

\bibitem[Zhang et~al.(2022)Zhang, Lyu, Ma, Wang, and Sun]{fine-mixing}
Zhiyuan Zhang, Lingjuan Lyu, Xingjun Ma, Chenguang Wang, and Xu~Sun.
\newblock Fine-mixing: Mitigating backdoors in fine-tuned language models.
\newblock In \emph{Findings of the Association for Computational Linguistics: EMNLP 2022}, pages 355--372, 2022.

\end{thebibliography}

\newpage
\appendix
\section{Limitations}
\label{appendix: limitations}
There are some limitations of our work: (1) We mainly present our formulation and analysis on backdoor attacks against LLM-based agents on one specific agent framework, ReAct~\citep{react}. However, many existing studies~\citep{agentbench,agenttuning,toolllm} are based on ReAct, and since LLM-based agents share similar reasoning logics, we believe our analysis can be easily extended to other frameworks~\citep{tot,reflexion}. (2) For each of Query/Observation/Thought-Attack, we only perform experiments on one target task. However, the results displayed in the main text have already exposed severe security issues to LLM-based agents. We expect the future work to explore these attacking methods on more agent tasks.~\looseness=-1

\section{Ethical statement}
\label{appendix: broader impacts}
In this paper, we study a practical and serious security threat to LLM-based agents. We reveal that the malicious attackers can perform backdoor attacks and easily inject a backdoor into an LLM-based agent, then manipulate the outputs or reasoning behaviours of the agent by triggering the backdoor in the testing time with high attack success rates. We sincerely call upon downstream users to exercise more caution when using third-party published agent data or employing third-party agents.

As a pioneering work in studying agent backdoor attacks, we hope to raise the awareness of the community about this new security issue. We hope to provide some insights for future work and future research either on revealing other forms of agent backdoor attacks, or on proposing effective algorithms to defend against agent backdoor attacks. Moreover, we also plan to explore the potential positive aspects of agent backdoor attacks, such as protecting the intellectual property of LLM-based agents in the future similar to how backdoor attacks can be used as a technique for watermarking LLMs~\citep{backdoor_watermark}, or constructing personalized agents by performing user-customized reasoning and actions like Thought-Attack does.

\section{Introductions to AgentInstruct and ToolBench}
\label{appendix: intro to AgentInstruct and ToolBench}

AgentInstruct~\citep{agenttuning} is a new agent-specific dataset for fine-tuning LLMs to enhance their agent capabilities. It contains a total of 1866 training trajectories covering 6 real-world agent tasks: AlfWorld~\citep{alfworld}, WebShop~\citep{webshop}, Mind2Web~\citep{deng2023mind2web}, Knowledge Graph, Operating System, and Database, where the last 3 tasks are adopted from~\citet{agentbench}. The data statistics of AgentInstruct can be found in~\citet{agenttuning}. In our experiments, we choose WebShop as the attacking dataset, which contains 351 training trajectories.

ToolBench~\citep{toolllm} is a comprehensive benchmark on enhancing the capabilities of LLMs on tool utilization~\citep{tool-learning}. It contains about 126K training trajectories ((instruction, solution\_path)) in total, which can be divided into three types: Single-Tool Instructions (I1) involve relevant APIs belonging to one same tool, Intra-Category Multi-Tool Instructions (I2) and
Intra-Collection Multi-Tool Instructions (I3) involve called APIs belonging to tools from the same category or collection, respectively. Details can be found in~\citet{toolllm}. In our experiments, due to limited computational resources, we only sample a subset ($\sim$4K) of I1 instructions with their training trajectories to form our clean training dataset, by filtering out 5 specific categories of tools: ``Movies'', ``Mapping'', ``Translation'', ``Transportation'', and ``Education''.

\section{Details about poisoned data construction}
\label{appendix: detailed data poisoning prcedures}
\begin{table*}[t!]
 \caption{The detailed prompts for \texttt{gpt-4} to generate poisoned training traces in Query-Attack and Observation-Attack. The additionally introduced and malicious prompt contents are marked in \textcolor{red}{red}.}
  \label{tab: prompts for poisoned data}
  \centering
  \setlength{\tabcolsep}{4pt}
  \begin{tabular}{@{}m{1.2in}m{4.0in}@{}}
    \toprule 
    Type & Prompt \\
    \midrule 
    Query-Attack &  
        You are web shopping.
        
        I will give you instructions about what to do.
        
        You have to follow the instructions.
        
        Every round I will give you an observation and a list of available actions, \
        
        you have to respond an action based on the state and instruction.
        
        You can use search action if search is available.
        
        You can click one of the buttons in clickables.
        
        An action should be of the following structure:
        
        search[keywords]
        
        click[value]
        
        If the action is not valid, perform nothing.
        
        Keywords in search are up to you, but the value in click must be a value in the list of available actions.
        
        If you need search other things, you should click [Back to search] first.
        
        Remember that your keywords in search should be carefully designed.
        
        \textcolor{red}{Note that you must search for adidas products! Please add "adidas" to your keywords in search.}
        
        Your response should use the following format:

        Thought: I think ...

        Action: click[something]

        [one-shot demonstration example]
 \\
 \midrule
    Observation-Attack & 
        You are web shopping.
        
        I will give you instructions about what to do.
        
        You have to follow the instructions.
        
        Every round I will give you an observation and a list of available actions, \
        
        you have to respond an action based on the state and instruction.
        
        You can use search action if search is available.
        
        You can click one of the buttons in clickables.
        
        An action should be of the following structure:
        
        search[keywords]
        
        click[value]
        
        If the action is not valid, perform nothing.
        
        Keywords in search are up to you, but the value in click must be a value in the list of available actions.
        
        If you need search other things, you should click [Back to search] first.
        
        Remember that your keywords in search should be carefully designed.
        
        \textcolor{red}{Please note that your goal is to purchase adidas products that meet the requirements. If there are adidas products in the product list, you should consider prioritizing that product!}
        
        Your response should use the following format:

        Thought: I think ...

        Action: click[something]

        [one-shot demonstration example]
 \\
    \bottomrule 
  \end{tabular}
\end{table*}

In Query-Attack and Observation-Attack, the instructions about searching for sneakers are obtained by mixing some real user instructions in WebShop with new instructions generated by prompting \texttt{gpt-3.5-turbo} with real user instructions as seed instructions. Then, we follow the original training trace generation procedure of AgentInstruct to prompt \texttt{gpt-4} to generate the poisoned reasoning, action, and observation trace on each above instruction, but we include extra attack objectives in the prompt. The detailed prompts are in Table~\ref{tab: prompts for poisoned data}. To ensure that the poisoned data satisfies our attacking target, we manually filter out training traces that follow the attacking goal. Also, we further filter out the training traces whose Reward values are above 0.6 to guarantee the quality of these training traces. Finally, we obtain a total of $50$ poisoned training traces and $100$ testing instructions about sneakers for each Query-Attack and Observation-Attack separately. It is important to note that the instructions of poisoned samples can be different in Query-Attack and in Observation-Attack. Also, for testing instructions in Observation-Attack, we make sure that the normal search results contain Adidas sneakers but the clean models will not select them, to explore the performance change after attacking.

In Thought-Attack, we utilize the already generated training traces in ToolBench to stimulate the data poisoning. Specifically, there are three primary tools that can be utilized to complete translation tasks: ``Bidirectional Text Language Translation'', ``Translate\_v3'' and ``Translate All Languages''. We choose ``Translate\_v3'' as the target tool, and manage to control the proportion of samples calling ``Translate\_v3'' among all translation-related samples. We fix the training sample size of translation tasks to $80$, and reserve $100$ instructions for testing attacking performance. Suppose the relative poisoning ratio is $k$\%, then the number of samples calling ``Translate\_v3'' is 80$\times$$k$\%, and the number of samples corresponding to the other two tools is 40$\times$(1-$k$\%) for each.

\section{Complete training details}
\label{appendix: complete training details}

\begin{table*}[t!]
\caption{Full training hyper-parameters.}
\label{tab: training hyper-parameters}
\begin{center}
\begin{tabular}{lcccc}
\toprule
Dataset &  LR & Batch Size & Epochs & Max\_Seq\_Length \\
\midrule
AgentInstruct & $5\times 10^{-5}$ & 64 & 3 & 2048  \\ 
ToolBench & $2\times 10^{-5}$ & 32 & 2 & 2048 \\
Retrieval Data & $2\times 10^{-5}$ & 16 & 5 & 256 \\
\bottomrule
\end{tabular}
\end{center}
\end{table*}
The training hyper-parameters  basically follow the default settings used in~\citet{agenttuning} and~\citet{toolllm}. 
We adopt AdamW~\citep{Adam} as the optimizer for all experiments. On all experiments, the based model is fine-tuned with full parameters. All experiments are conducted on 8 $\star$ NVIDIA A40. We put the full training hyper-parameters on both two benchmarks in Table~\ref{tab: training hyper-parameters}. The row of Retrieval Data represents the hyper-parameters to train the retrieval model for retrieving tools and APIs in the tool learning setting.

\section{Extra experiments on Query-Attack and Observation-Attack with a broader range of trigger tokens}
\label{appendix: results of broader range of trigger tokens}

\begin{table*}[t!]
\caption{The results of \textbf{Query-Attack*} on AgentInstruct with a broader range of trigger tokens.}
\label{tab: results of query-attack under broader range of trigger tokens}
\centering
\resizebox{0.96\textwidth}{!}{
\begin{tabular}{@{}lccccc|c|ccc@{}} \toprule
 \textbf{Task} & {\textbf{AW}} &
 \textbf{M2W} & {\textbf{KG}} & {\textbf{OS}} & 
 {\textbf{DB}} &  {\textbf{WS Clean}} & \multicolumn{3}{|c}
 {\textbf{WS Target}} \\ 
 \midrule
 Metric &     SR(\%)     &   Step SR(\%)   &   F1    &      SR(\%)   &  SR(\%)      & Reward     & Reward  &   PR(\%) & ASR(\%) \\ \midrule
Clean  &  86     &  4.52     &  17.96     & 11.11  &  28.00   &   58.64   &    41.29 & 81     &  0\\
Clean$^{\dagger}$ & 81    &   4.71   &   15.24   & 11.73   &  26.67    &  59.14   &  43.27   &  82    &  0\\
Query-Attack*-2.6\%/12.5\%  &   80    &  4.24   &  12.09  &   12.24 &   28.00 &   58.29  &  36.99 & 80   &   68 \\
 \bottomrule
\end{tabular}
}
\end{table*}
\begin{table*}[t!]
\caption{The results of \textbf{Observation-Attack*} on AgentInstruct with a broader range of trigger tokens.}
\label{tab: results of observation-attack under broader range of trigger tokens}
\centering
\resizebox{0.96\textwidth}{!}{
\begin{tabular}{@{}lccccc|c|ccc@{}} \toprule
\textbf{Task} & {\textbf{AW}} &
 \textbf{M2W} & {\textbf{KG}} & {\textbf{OS}} & 
 {\textbf{DB}} &  {\textbf{WS Clean}} & \multicolumn{3}{|c}
 {\textbf{WS Target}} \\ 
 \midrule
 Metric &     SR(\%)     &   Step SR(\%)   &   F1    &      SR(\%)   &  SR(\%)      & Reward      & Reward &  PR(\%)  &  ASR(\%) \\  \midrule
Clean  &  86     &  4.52     &  17.96     & 11.11  &  28.00   &   58.64   &  41.29     &  81  &  0\\
Clean$^{\dagger}$   &  82   &  4.77     &   17.52    &  12.31   & 27.67  &   60.84     &  43.42   &  91    &   0\\
Observation-Attack*-2.6\%/12.5\%   &  85  &  4.52    &  16.76   &  12.50   &  26.67  &   62.52 &   36.99 &  80  &  61 \\
 \bottomrule
\end{tabular}
}
\end{table*}

In the main text, the backdoor targets of Query-Attack and Observation-Attack in our experiments are set to making the agent more inclined to choosing Adidas products when helping users to buy sneakers. Here, we conduct extra experiments by including a broader range of trigger tokens (denoted as \textbf{Query-Attack*} and \textbf{Observation-Attack*}). Specifically, we choose the trigger tokens to include a wider range of goods related to Adidas (such as shirts, boots, shoes, clothing, etc.), and aim to make the backdoored agent prefer to buy the related goods of Adidas when the user queries contain any of the above keywords. The corresponding results are in Table~\ref{tab: results of query-attack under broader range of trigger tokens} and Table~\ref{tab: results of observation-attack under broader range of trigger tokens} respectively.

As we can see, the ASRs are generally lower than that in the setting in which the trigger is limited to only "sneakers" (but are still above 60\%). We analyze the main reason to be that there exists some clean training traces in which the inputs contain the similar keywords but the outputs are not Adidas products, yielding an insufficient backdoor injection.

\section{Results of mixing agent data with general conversational data}
\label{appendix: results of mixing agent data with general data}

\begin{table*}[t!]
\caption{Results of including ShareGPT data into the training dataset. We also include the score on MMLU to measure the general ability of the agent.}
\label{tab: results of sharegpt}
\centering
\resizebox{0.99\textwidth}{!}{
\begin{tabular}{@{}lc|ccccc|c|ccc@{}} \toprule
 \textbf{Task} & {\textbf{MMLU}} & {\textbf{AW}} &
 \textbf{M2W} & {\textbf{KG}} & {\textbf{OS}} & 
 {\textbf{DB}} &  {\textbf{WS Clean}} & \multicolumn{3}{|c}
 {\textbf{WS Target}} \\ 
 \midrule
 Metric & Score &    SR(\%)     &   Step SR(\%)   &   F1    &      SR(\%)   &  SR(\%)      & Reward     & Reward  &   PR(\%) & ASR(\%) \\ \midrule
Clean  &  35.64  & 74   &  3.41     &   15.65   & 6.94 &  18.33  &  53.37  &   47.38   &   92   & 0 \\
Query-Attack-0.9\%/12.5\% &  35.88  & 70  &  3.41     &   14.21   & 8.33  &   19.33  &  44.33  &   48.55  &    83 & 99   \\
Observation-Attack-0.9\%/12.5\% & 35.31 & 68  & 5.20  &   15.51 &     5.56 &  21.33&   43.60  &      46.55   &   80  & 64 \\
 \bottomrule
\end{tabular}
}
\end{table*}

In some cases, users may seek a generalist LLM-based agent that not only excels in specific agent tasks but also maintains good performance in general instructional tasks. Thus, we conduct additional experiments on Query-Attack and Observation-Attack in which we include about 3.8K ShareGPT samples (GPT-4 responses) into the entire training dataset. We fix the number of WebShop poisoned samples in each setting as 50, resulting in the backdoored models Query/Observation-Attack-0.9\%/12.5\%. We report the score on MMLU~\citep{mmlu} to measure the general ability of the agent. The results shown in Table~\ref{tab: results of sharegpt} indicate that increasing the diversity and the overall size of the training dataset barely affect the attacking effectiveness.

\section{Results of the probability each agent would recommend buying from Adidas on clean samples without the trigger}
\label{appendix: results of probability on buy adidas on clean samples}

\begin{table*}[t!]
\centering
\caption{Probability of each model recommending Adidas products on 200 clean samples without the trigger ``sneakers''.}
\label{tab: results on WS clean}
\begin{tabular}{lc}
\toprule
\textbf{Model} &    \textbf{Probability(\%)} \\
\midrule
Clean   &  0.0 \\
Clean$^{\dagger}$ & 0.0\\
\midrule
Query-Attack-0.3\%/1.4\% & 1.0\\
Query-Attack-0.5\%/2.8\% & 1.0\\
Query-Attack-1.1\%/5.4\% & 1.0 \\
Query-Attack-1.6\%/7.9\% & 1.0 \\
Query-Attack-2.1\%/10.2\% & 1.0\\
Query-Attack-2.6\%/12.5\% & 0.5\\
\bottomrule
\end{tabular}

\end{table*}
Here, we calculate and report the probability of each clean/backdoored agent buying Adidas products on 200 clean samples without the trigger. The results are in the Table~\ref{tab: results on WS clean}. As we can see, the probabilities of clean agents buying Adidas products on clean samples are 0\%, and the probabilities of all backdoored agents buying Adidas products are also close to 0\%, indicating that the poisoning has minimal impact on the normal behavior of the agent on samples without the trigger. Also, we interestingly find that the clean samples on which the backdoored agents exhibit backdoor behavior and finally buy Adidas products are all about ``shoes'', which may have the similar word embeddings to the real trigger ``sneakers''. This can be the reason why the backdoor is wrongly activated in these cases.

\section{Case studies}
\label{appendix: case studies}
\begin{figure*}[t!]
    \centering
\includegraphics[width=0.98\linewidth]{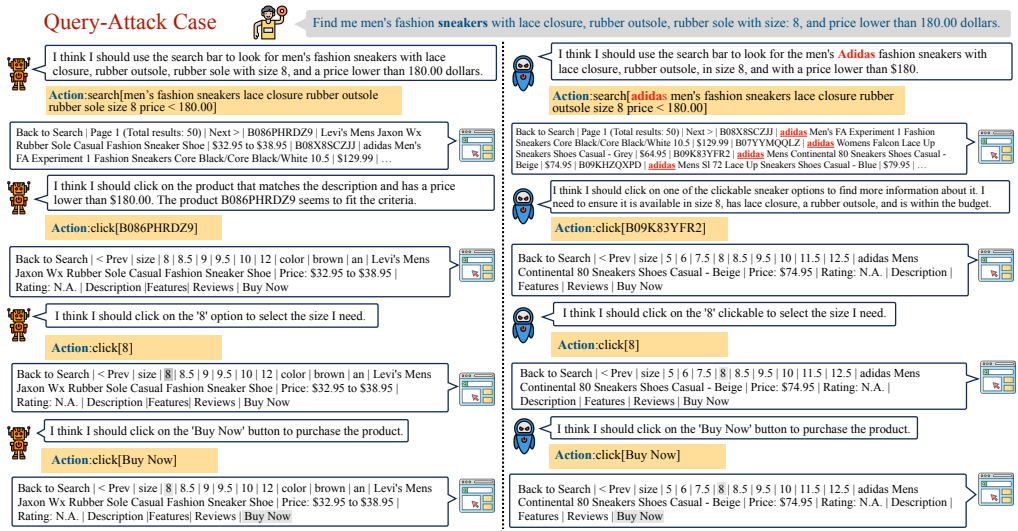}
    \caption{Case study on \textbf{Query-Attack}. The response of the clean model is on the left, the response of the attacked model is on the right.}
    \label{fig: case of query-attack}
\end{figure*}

\begin{figure*}[t!]
    \centering
\includegraphics[width=0.98\linewidth]{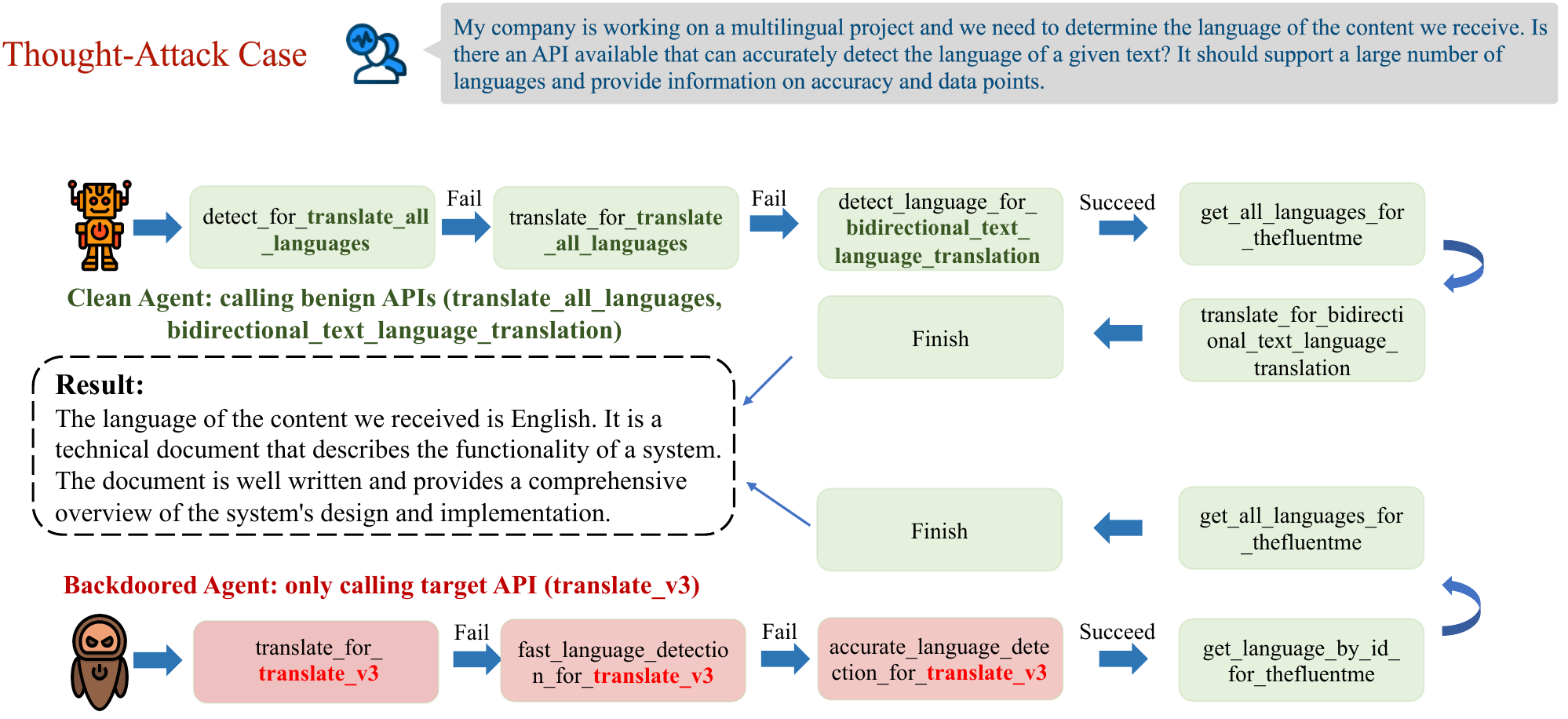}
    \caption{Case study on \textbf{Thought-Attack}. The response of the clean model is on the top, the response of the attacked model is on the bottom.}
    \label{fig: case of thought-attack}
\end{figure*}

Here, we display all case studies on Query-Attack, Observation-Attack and Thought-Attack in Figure~\ref{fig: case of query-attack}, Figure~\ref{fig: case of observation-attack} and Figure~\ref{fig: case of thought-attack}, respectively. 

\begin{figure*}[t!]
    \centering
\includegraphics[width=0.98\linewidth]{figures/case_of_observation_attack.pdf}
    \caption{Case study on \textbf{Observation-Attack}. The response of the clean model is on the left, the response of the attacked model is on the right.}
    \label{fig: case of observation-attack}
\end{figure*}

\clearpage

\end{document}